\documentclass[letterpaper,draftcls,onecolumn,journal,11pt]{IEEEtran}

\usepackage[dvips]{graphicx}
\usepackage{psfrag}
\usepackage{citesort}

\usepackage{amsmath,amssymb,amsfonts}

\usepackage{pifont}
\renewcommand{\baselinestretch}{1.0}

\date{}

\normalfont

\newtheorem{theorem}{Theorem}
\newtheorem{lemma}{Lemma}

\begin{document}

\title{Collaborative 
Beamforming for Distributed Wireless Ad Hoc Sensor Networks
\footnote{%
This material is based upon research supported in part by
the National Science Foundation under the Alan T. Waterman
Award, Grant No. CCR-0139398, in part by the Office of
Naval Research under Grant No. N00014-03-1-0102, and in
part by a Fellowship from the John Simon Guggenheim
Memorial Foundation.  Any opinions, findings, and conclusions
or recommendations expressed in this publication are
those of the authors and do not necessarily reflect the
views of the National Science Foundation. \\  \quad
This paper was presented in part
at the IEEE Information Theory Workshop (ITW'04),
San Antonio, TX, October 2004, and in part
at the IEEE International Conference on Acoustics,
Speech, and Signal Processing (ICASSP'05),
Philadelphia, PA, March 2005. 
}}
\author{\authorblockN{Hideki Ochiai} \\
\authorblockA{Department of Electrical 
and Computer Engineering, Yokohama National University}\\
\and
\authorblockN{Patrick Mitran} \\
\authorblockA{Division of Engineering 
and Applied Sciences, Harvard University}\\
\and
\authorblockN{H. Vincent Poor} \\ 
\authorblockA{Department of Electrical Engineering,
Princeton University} \\
\and
\authorblockN{Vahid Tarokh} \\
\authorblockA{Division of Engineering and Applied Sciences, 
Harvard University}\\
}
\date{}
\maketitle

\begin{abstract}
The performance of collaborative beamforming
is analyzed using the theory of random arrays. 
The statistical 
average and distribution of the beampattern of randomly 
generated phased arrays is derived in the framework of 
wireless ad hoc sensor networks. Each sensor node 
is assumed to have a single isotropic antenna and
nodes in the cluster collaboratively transmit the signal
such that the signal in the target direction 
is coherently added in the far-field region.
It is shown that with $N$ sensor nodes uniformly distributed
over a disk, the directivity can approach $N$, provided
that the nodes are located sparsely enough. The distribution of 
the maximum sidelobe peak is also studied.
With the application to ad hoc networks in mind, 
two scenarios, closed-loop and open-loop, 
are considered. Associated with these scenarios,
the effects of phase jitter and location estimation errors 
on the average beampattern are also analyzed.
\end{abstract}

\begin{center}
{\em To Appear in IEEE Transactions on Signal Processing, 2005.}
\end{center}

\section{Introduction}

Recent advances in the construction of low cost, low power, and mass
produced micro sensors and 
Micro-Electro-Mechanical (MEM) systems 
have ushered in a new
era in system design using 
distributed sensor networks \cite{Yao:1998,Chen:2003}.
The advent of sensor network technology provides a variety of applications
that have been considered unrealistic in the past. 
One such application is in the area of space communications:
with ad hoc sensor networks,
a number of sensor nodes randomly placed on a planet 
can collaboratively collect information 
and then, also collaboratively, send the information back to
Earth. In this scenario, the sensors must have an ability to
transmit information over very long distances 
with high energy efficiency. In this kind of point-to-point
communication, directional antennas are a preferred means 
to avoid interference.

In general, this can be achieved by adaptive beamforming.
Given a number of well-designed antenna elements at the 
transmitting/receiving sensor nodes, each node could 
in principle autonomously  transmit/receive
the information to/from any desired direction.
The advantages and applications of beamforming with antenna
arrays are well known; in wireless communications, 
this enables Space-Division Multiplex Access (SDMA), a
technology which has the potential to significantly increase 
the capacity of multiple access channels.

\begin{figure}[t]
\begin{center}
\includegraphics[width=11.5cm,clip]{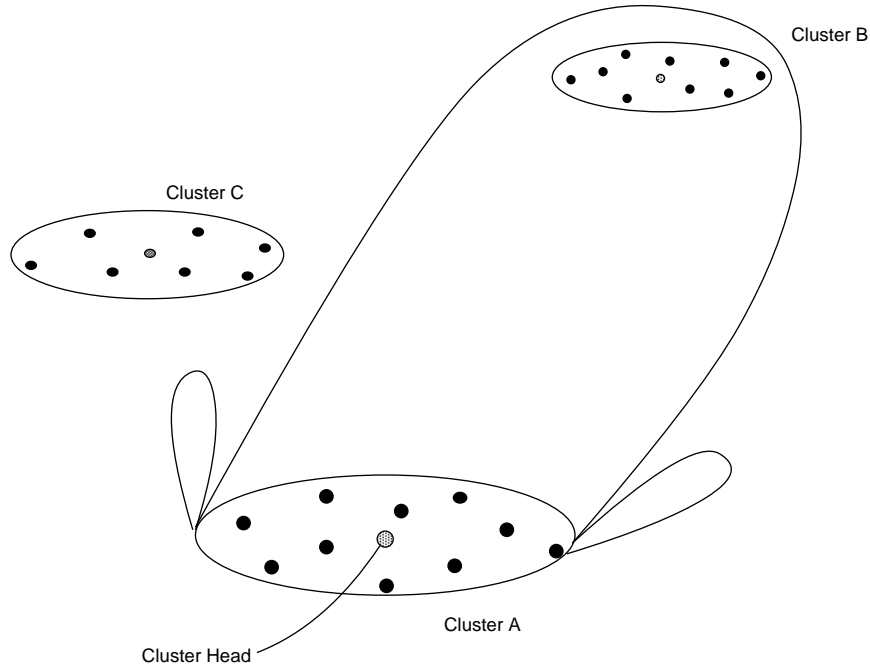}
\caption{Collaborative beamforming concept in ad hoc sensor networks.}
\label{fig:adhoc_network_model}
\end{center}
\vspace{-1cm}
\end{figure}

One of the most important constraints on wireless sensors 
is energy efficiency. Since the sensor nodes are 
often distributed 
in places where manual maintenance is costly, such as 
remote locations, on top of buildings and so on, 
it should be possible to operate these for
several months without battery replacement.
Considering the fact that each antenna element requires 
analog circuitry (and thus leads to costly hardware),
in practice each distributed sensor is likely to be
equipped with only a single antenna 
and this precludes the use of autonomous
beamforming in scenarios of very energy efficient communication.
Nevertheless, if the sensors in the cluster share the information
{\em a priori} and synchronously transmit 
the data collaboratively as sketched in 
Fig.\,\ref{fig:adhoc_network_model},
it may be possible to beamform 
when transmitting (and also receiving) the data
in a distributed manner. 
The resultant overhead due to intra-cluster information sharing
may be relatively small as this can be done by low-cost
short distance broadcasting-type communication among nodes.
Thus, with distributed collaborative beamforming, the nodes can 
send the collected information to the far-end receiver
over long distances with high efficiency. 
Also, only the sensor cluster in the specified target
direction receives the data with high signal power
and no significant interference occurs for clusters
in other directions. 
Overall there is thus a potential to 
increase the capacity of the multiple access channel significantly
despite the additional overhead for information sharing.

The obvious question is whether
one can form a nice beampattern with a narrow mainbeam, 
or achieve a reasonable directional gain (directivity).
The sensor nodes in ad hoc networks are located randomly
by nature, and the resultant beampattern
depends on the particular realization of the sensor node 
locations. Therefore, it may be quite natural to treat the
beampattern with probabilistic arguments. 
In this paper, assuming idealized 
channel model conditions and antenna properties,
we analyze the achievable performance of 
collaborative beamforming based on distributed sensor nodes
in a probabilistic sense. 
Specifically, the statistical properties of the achievable 
beampattern of the random sensor arrays are studied based
on the following assumptions. The sensors form an ad hoc network 
and the geometry of the cluster is given by a two-dimensional
disk of a given radius over which all sensor nodes are distributed 
uniformly as illustrated in Fig.\,\ref{fig:adhoc_network_model}.
Since the corresponding far-field
beampattern depends on the particular realization of the
random array of nodes, 
the probability distribution of the far-field beampattern
is of particular interest. 

To the best of the authors' knowledge, the beampattern 
aspects of collaborative beamforming using random arrays have 
not been analyzed before in the framework of wireless ad hoc 
sensor networks.
Nevertheless, in the antenna design literature, probabilistic 
analysis of random arrays is not new. 
In the framework of linear array design with a large number of sensors,
Lo \cite{Lo:1967} has developed a comprehensive theory of random arrays
in the late 1960s. It has been shown that
randomly generated linear arrays with a large number of nodes
can in fact form a good beampattern with high probability\footnote{%
It is interesting to note that 
the theory of random arrays has been discussed and developed
almost exclusively in the antenna design community, e.g.,
in \cite{Lo:1967,Steinberg:1972,Agrawal:1972,Donvito:1979}. 
} and that with linear random arrays of $N$ 
sensors, the directivity approaches $N$ asymptotically. 
Although our scenario is quite different in that our main
goal is not to design array geometry but to {\em exploit}
the randomness of the distributed sensor network,
it turns out that the results we shall develop in this paper 
can be seen as an extension of the theory of linear random arrays, 
\cite{Lo:1967}, to random arrays on a disk. 
Thus, the same conclusion will be reached:
with $N$ collaborative sensor nodes, one can achieve
a directivity of order $N$ asymptotically. 

The major difference between classical beamforming by antenna arrays
and distributed beamforming is that whereas 
the geometry of the former
is usually known {\em a priori},
the exact location of the sensor 
nodes in ad hoc network is not, and it should be acquired dynamically. 
Even if their relative location is estimated by some 
adaptive algorithm (e.g., \cite{Yao:1998} for receiver beamforming), 
considering the low SNR operation of the sensor nodes, 
it is almost certain that
the acquired geometric information has some inaccuracy. 
Also, since all nodes are operated with physically
different local oscillators,
each node may suffer from statistically independent phase offsets.
In order to model and elucidate the effect of these impairments,
we consider the following two scenarios: {\em closed-loop}
and {\em open-loop}. The closed-loop scenario
may be described as follows. 
Each node independently synchronizes itself to the beacon sent 
from the destination node (such as a base station)
and adjusts its initial phase to it.
Thus, the beam will be formed in the direction of arrival of the beacon.
This kind of system is often referred to as a {\em self-phasing} array
in the literature, and may be effective for systems operating 
in Time-Division Duplex (TDD) mode. The residual phase jitter due
to synchronization and phase offset estimation among sensor nodes
is then often the dominant impairment. 
On the other hand, in the open-loop scenario we assume
that all nodes within the cluster acquire their 
relative locations from the beacon of a nearby reference point or cluster head.
The beam will then be steered toward an arbitrary direction.
Thus, the destination need not transmit a beacon,
but each node requires knowledge of its relative
position from a predetermined reference point within the cluster.
This case may occur
in ad hoc sensor networks where sensor nodes do not
have sufficient knowledge of the
destination direction {\em a priori}. 
In this scenario, since the acquisition of precise knowledge 
is not realistic, the effects of location estimation ambiguity among
sensors upon the beampattern may be of particular interest.

Throughout the paper, the nodes and channel are assumed to be static 
over the communication period, and 
for simplicity the information rate is assumed to be sufficiently low 
that Inter-Symbol Interference (ISI), due to residual
timing offset, is negligible.
It will also be assumed that all nodes share the same transmitting
information {\em a priori}, as the main focus of the paper is on the
beampattern, rather than the front-end communication performance.

The paper is organized as follows. 
Section \ref{sec:model} describes
the assumptions, model, and main parameters that describe beam
characteristics associated with the framework of wireless ad
hoc sensor networks. 
In Section 
\ref{sec:average_property},
the average properties of the beampattern are derived.
The average beampattern
of linear random arrays
has been derived in \cite{Lo:1967}, 
and our results can be seen as its extension 
to our sensor network model. 
For analytical purpose, we also introduce 
the concept of 3 dB sidelobe region.
In Section \ref{sec:distribution_beam}, 
the statistical distribution of
the beampattern in a specific direction is derived.
Lo \cite{Lo:1967} has derived the distribution of
the beampattern in linear arrays
based on a Gaussian approximation of the array factor,
which is a common assumption in the random array literature.
In contrast, we shall develop a numerical method to calculate the exact
distribution of the beampattern and also examine
the accuracy of the Gaussian approximation in detail. 

The distribution of the maximum of the sidelobe region
is discussed in Section 
\ref{sec:distribution_maximum}. This aspect of beampattern
was analyzed by Steinberg \cite{Steinberg:1972}, 
Agrawal and Lo \cite{Agrawal:1972}, and 
Donvito and Kassam \cite{Donvito:1979} in the framework
of linear random arrays. In this paper, we derive
an upper bound on the distribution of the maximum sidelobe
in our framework of collaborative beamforming
based on the approach of \cite{Donvito:1979}.
The effect of phase jitter or location estimation errors 
on 
the resultant beampattern,
associated with the closed-loop and open-loop scenarios,
is analyzed in Section \ref{sec:imperfect}.
Steinberg \cite{Steinberg:1976} has analyzed the effects of
phase estimation errors in linear arrays, and 
based on a similar approach we analyze the effects of the
average beampattern with phase estimation errors.
Finally, Section \ref{sec:conclusion} concludes the paper.

\section{System Model and Beampattern}
\label{sec:model}

\begin{figure}[t]
\begin{center}
\psfrag{z}[Bl][Bl][.8]{$z$}
\psfrag{x}[Bl][Bl][.8]{$x$}
\psfrag{y}[Bl][Bl][.8]{$y$}
\psfrag{R}[Bl][Bl][.8]{$R$}
\psfrag{r1phi1}[Bl][Bl][.8]{$(r_1, \psi_1)$}
\psfrag{rkphik}[Bl][Bl][.8]{$(r_k, \psi_k)$}
\psfrag{Aphi0theta0}[Bl][Bl][.8]{$(A, \phi_0, \theta_0)$}
\psfrag{theta}[Bl][Bl][.8]{$\theta_0$}
\psfrag{phi}[Bl][Bl][.8]{$\phi_0$}
\includegraphics[width=12.5cm,clip]{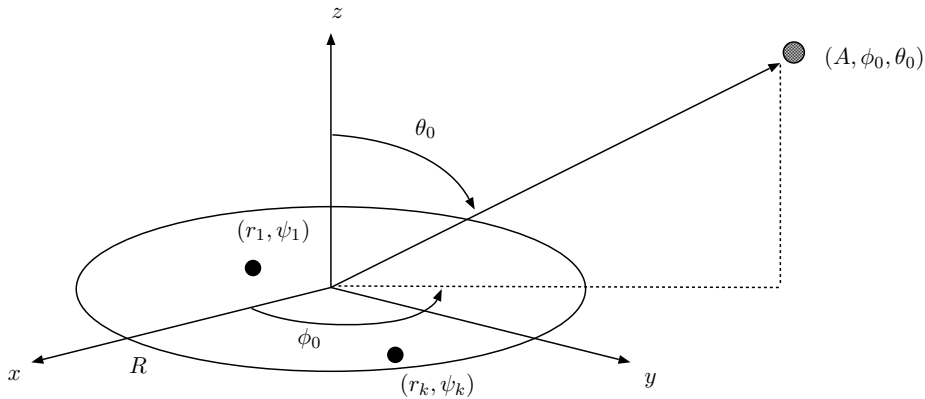}
\caption{Definitions of notation.}
\label{fig:notaion}
\end{center}
\vspace{-1cm}
\end{figure}

The geometrical configuration of the distributed nodes and 
destination (or target) is illustrated in Fig.\,\ref{fig:notaion}
where, without loss of generality,
all the collaborative sensor nodes are assumed to be located on the
$x$-$y$ plane. The $k$th node location is thus denoted 
in polar coordinates by $(r_k, \psi_k)$. 
The location of the destination 
is given in spherical coordinates
by $(A, \phi_0, \theta_0)$. Following
the standard notation in antenna theory \cite{Balanis:1997}, 
the angle $\theta \in [0, \pi]$
denotes the elevation direction, whereas the angle $\phi \in [-\pi, \pi]$ 
represents the azimuth direction.
In order to simplify the analysis, the following assumptions are
made: 
\begin{enumerate}
\item The location of each node is chosen randomly, following 
a uniform distribution within a disk of radius $R$.
\item Each node is equipped with a single ideal isotropic antenna.
\item All sensor nodes transmit identical energies, and the path losses
of all nodes are also identical. Thus the underlying model
falls within the framework of phased arrays.
\item There is no reflection or scattering of the signal.
Thus, there is no multipath fading or shadowing.
\item 
The nodes are sufficiently separated
that any mutual coupling effects \cite{Balanis:1997} 
among the antennas of different sensor nodes are negligible.
\end{enumerate}

Furthermore, we also assume that all the nodes are perfectly
synchronized so that no frequency offset or phase jitter occurs. 
The effects of phase ambiguities on the beampattern 
will be discussed in Section\,\ref{sec:imperfect}.

Let $d_k ( \phi, \theta )$ denote the Euclidean distance between 
the $k$th node and the reference location $(A, \phi, \theta)$, 
which is written as
\begin{align}
d_k ( \phi, \theta )
& = \sqrt{
 A^2 + r^2_k - 2 r_k  A \sin \theta \cos( \phi - \psi_k )
} \,\,.
\label{dist0}
\end{align}
If the initial phase of node $k \in \{1,2,\ldots, N\}$ 
is set to 
\begin{equation}
\Psi_k = - \frac{2\pi}{\lambda} d_k(\phi_0, \theta_0 ),
\label{varphik1}
\end{equation}
the corresponding array factor, given the realization
of node locations ${\boldsymbol r} = [r_1, r_2, \ldots, r_N] \in
[0, R]^N$
and ${\boldsymbol \psi} = [\psi_1, \psi_2, \ldots, \psi_N] \in [-\pi, \pi]^N$,
is written as
\begin{align}
F(\phi, \theta | {\boldsymbol r},{\boldsymbol \psi} ) 
& = \frac{1}{N}
\sum_{k=1}^{N} 
e^{j \Psi_k}
e^{j \frac{2 \pi}{\lambda}
d_k(\phi, \theta)} 
= \frac{1}{N}
\sum_{k=1}^{N} e^{j \frac{2 \pi}{\lambda}
\left[
d_k(\phi, \theta) - d_k(\phi_0, \theta_0) \right]} ,
\label{far_field_F0}
\end{align}
where $N$ is the number of sensor nodes and
$\lambda$ is the wavelength of the radio frequency (RF) carrier.

In this paper, we are interested in the radiation
pattern in the far-field region, and we assume 
that the far-field condition $A \gg r_k$ holds.
The far-field distance 
$d_k ( \phi, \theta )$
in \eqref{dist0}
can then be approximated as 
\begin{align}
d_k ( \phi, \theta ) 
& \approx A - r_k \sin \theta \cos( \phi - \psi_k ).
\label{Dk_0}
\end{align}
The far-field beampattern is thus approximated by
\begin{align}
F\left( \phi, \theta  |
{\boldsymbol r},{\boldsymbol \psi} \right) 
& \approx \frac{1}{N}
\sum_{k=1}^{N} e^{j \frac{2 \pi}{\lambda}
r_k \left[ \sin \theta_0 \cos(\phi_0 - \psi_k) -  
\sin \theta \cos(\phi - \psi_k) \right]} 
\triangleq \tilde{F}(\phi, \theta | {\boldsymbol r},{\boldsymbol \psi} )  .
\label{far_field_F1}
\end{align}

Alternatively, instead of applying $\Psi_k$ as in \eqref{varphik1},
if we choose
\begin{equation}
\Psi_k^\dag = \frac{2\pi}{\lambda} r_k \sin \theta_0 \cos(\phi_0 - \psi_k ),
\label{varphik2}
\end{equation}
then we obtain the array factor as
\begin{align}
F^\dag (\phi, \theta | {\boldsymbol r},{\boldsymbol \psi} )  
& = \frac{1}{N} \sum_{k=1}^{N} 
e^{j \Psi^{\dag}_k}
e^{j \frac{2 \pi}{\lambda}
d_k(\phi, \theta)} \nonumber \\
& \approx \frac{1}{N}
\sum_{k=1}^{N} e^{j \frac{2 \pi}{\lambda}
\left[
A - r_k \sin \theta \cos( \phi - \psi_k )
+ r_k \sin \theta_0 \cos(\phi_0 - \psi_k )
\right]} 
\nonumber \\
& =
e^{j \frac{2 \pi}{\lambda} A }
\frac{1}{N}  \sum_{k=1}^{N} 
e^{j \frac{2 \pi}{\lambda}
r_k \left[ \sin \theta_0 \cos(\phi_0 - \psi_k) - 
\sin \theta \cos(\phi - \psi_k) \right]} 
 \triangleq \tilde{F}^{\dag} (\phi, \theta
| {\boldsymbol r},{\boldsymbol \psi} )  .
\label{far_field_F2}
\end{align}
The only difference between $
\tilde{F}(\phi, \theta | {\boldsymbol r},{\boldsymbol \psi} )$ 
in \eqref{far_field_F1} and 
$\tilde{F}^{\dag}(\phi, \theta
| {\boldsymbol r},{\boldsymbol \psi} )$  
in \eqref{far_field_F2} is 
the existence of the initial phase offset
of $\frac{2\pi}{\lambda} A$.
The far-field beampattern is thus identical for both systems,
and the received signal exhibits no difference
as long as the base station compensates for 
this phase rotation.

Therefore, there are two ways of forming a beam.
One way is to use \eqref{varphik1}, but this approach requires
accurate knowledge of the distance, relative to the wavelength $\lambda$,
between each node and the destination.
Thus, this applies to the closed-loop case such as
self-phasing arrays. 
Alternatively, the use of \eqref{varphik2} requires
knowledge of the node positions relative to some
common reference (such as the origin in this example), and
thus corresponds to the open-loop case.
Knowledge of the elevation direction $\theta_0$ is also required, but
this may be assumed to be known {\em a priori} in many applications.
In both cases, the synchronization among sensors is critical, which may
be achieved by the use of reference signals such as those of the Global
Positioning System (GPS).

Of particular interest in practice is 
the case where $\theta_0 = \frac{\pi}{2}$,
i.e., the destination node is in the same plane 
as the collaborative sensor nodes.
Therefore,  
we will consider the beampattern in this plane and thus
assume that $\theta = \theta_0 = \frac{\pi}{2}$ 
for the rest of the paper.
We then denote 
$\tilde{F}(\phi, \theta = \pi/2 | {\boldsymbol r},{\boldsymbol \psi} )$
in \eqref{far_field_F1}
by $\tilde{F}(\phi | {\boldsymbol r},{\boldsymbol \psi} )$ 
and 
$\tilde{F}^{\dag}(\phi, \theta = \pi/2
| {\boldsymbol r},{\boldsymbol \psi} )$  
in \eqref{far_field_F2} by
$\tilde{F}^{\dag}(\phi | {\boldsymbol r},{\boldsymbol \psi} )$
for simplicity.

By assumption, the node locations $(r_k, \psi_k)$
follow the uniform distribution over the disk of radius
$R$. Thus, the probability density functions (pdfs) of $r_k$ and $\psi_k$
are given by
\begin{align}
f_{r_k}(r)  = \frac{2 r}{R^2}, \quad 0 < r < R, 
\quad
\mbox{ and }
\quad 
f_{\psi_k}(\psi)  = \frac{1}{2 \pi}, \quad -\pi  \leq \psi < \pi  .
\nonumber 
\end{align}
From \eqref{far_field_F1}, we have (with $\theta = \theta_0 = \pi/2$)
\begin{align}
\tilde{F}(\phi | {\boldsymbol r},{\boldsymbol \psi} )    
& = 
\frac{1}{N} \sum_{k=1}^{N} e^{j \frac{4 \pi}{\lambda}
r_k \sin \left( \frac{\phi_0 - \phi}{2} \right)
\sin \left(
\psi_k - \frac{\phi_0 + \phi}{2}
\right)} 
=
\frac{1}{N} \sum_{k=1}^{N} e^{j {4 \pi} \frac{R}{\lambda}
\sin \left( \frac{\phi_0 - \phi}{2} \right)
\tilde{r}_k
\sin 
\tilde{\psi}_k },
\label{far_field_F3}
\end{align}
where $\tilde{r}_k \triangleq r_k / R$ and
$\tilde{\psi}_k \triangleq \psi_k - \frac{\phi_0 + \phi}{2}$.
The compound random variable
\begin{equation}
z_k \triangleq 
\tilde{r}_k
\sin \tilde{\psi}_k ,
\end{equation}
has the following pdf:
\begin{equation}
f_{z_k} (z) = \frac{2}{\pi} \sqrt{1 - z^2}, \quad -1 \leq z \leq 1 .
\label{pdf_z}
\end{equation}
Note that since the above model is symmetric with respect to 
the azimuth direction $\phi$, any particular choice of
$\phi_0$ does not change the results in the following. 
Therefore, without loss of generality, we assume that $\phi_0 =0$,
and the parameter $\phi$ simply corresponds to the difference angle
between the target direction and the reference.
Also, note that $|\phi| \leq \pi$.

The array factor of
\eqref{far_field_F3} can then be rewritten as
\begin{align}
\tilde{F}(  \phi | {\boldsymbol z} )
& =
\frac{1}{N} \sum_{k=1}^{N} e^{-j {4 \pi} \tilde{R}
\sin \left( \frac{ \phi}{2} \right)
z_k}   ,
\label{far_field_F4}
\end{align}
where 
$\tilde{R} \triangleq \frac{R}{\lambda}$ is the radius
of the disk normalized by the wavelength.

Finally, the far-field beampattern can be defined as 
\begin{align}
P(\phi | {\boldsymbol z} ) &
\triangleq
\left| 
\tilde{F}(\phi | {\boldsymbol z })
 \right|^2 =
\tilde{F}(\phi | {\boldsymbol z })
\tilde{F}^* (\phi | {\boldsymbol z } ) 
\nonumber \\
&  =
 \frac{1}{N^2} \sum_{k=1}^{N}
\sum_{l=1}^{N}
e^{ - j {4 \pi} \tilde{R}
\sin \left( \frac{\phi}{2} \right)
(z_k - z_l)} 
\nonumber \\
&  =\frac{1}{N} +
\frac{1}{N^2} \sum_{k=1}^N 
e^{ - j \alpha (\phi) z_k}
\sum_{\stackrel{ \scriptstyle l=1}{l\neq k}}^{N} 
e^{ j \alpha (\phi) z_l}
\label{beampattern1}
\end{align}
where 
\begin{align}
\alpha(\phi) & \triangleq {4 \pi} \tilde{R} \sin  \frac{ \phi}{2} .
\label{alpha_def}
\end{align}

\section{Average Properties of Beampattern of
Uniformly Distributed Sensor Array with Perfect Phase Information}
\label{sec:average_property}

\subsection{Average Far-Field Beampattern}

We start by investigating the average beampattern
of the random array resulting from the distributed sensor 
network model in the previous
section. Here, the average is taken over all realizations of
${\boldsymbol z}$, and from \eqref{beampattern1}
the average beampattern is expressed as
\begin{align}
P_{\text{av}}( \phi )
& \triangleq E_{\boldsymbol z}\left\{
P( \phi | {\boldsymbol z} ) 
\right\},
\label{av_beam0}
\end{align}
where $E_{\boldsymbol x}\{\cdot\}$ denotes expectation 
with respect to the random variables ${\boldsymbol x}$.
From \eqref{beampattern1} and \eqref{pdf_z}, it can be readily shown
that
\begin{align}
P_{\text{av}}(  \phi )
& = \frac{1}{N} + \left( 1 - \frac{1}{N} \right)
\left| 2 \cdot \frac{J_1\left(
\alpha(\phi)  \right)}
{\alpha(\phi)}
\right|^2 ,
\label{av_beam1}
\end{align}
where $J_n(x)$ is the $n$th order 
Bessel function of the first kind. 
Although the function $J_1(x)/x$ is oscillatory, 
the local maxima of oscillation tend to decrease with 
increasing $x$. In \eqref{av_beam1}, the first
term represents the average power level of the sidelobe,
which does not depend on the node location,
whereas the second term is the contribution of the mainlobe
factor. Since, conditioned on $\phi$,
the array factor of the form \eqref{far_field_F4}
is an average of bounded independent and identically distributed
(i.i.d.) complex random variables,
by the weak law of large numbers the beampattern \eqref{beampattern1}
converges to the ensemble average 
\eqref{av_beam1} in probability as $N\to \infty$.

\begin{figure}[t]
\begin{center}
\psfrag{Power[dB]}[Bl][Bl][.8]{{\sf Average Power [dB]}}
\psfrag{Angle[degree]}[Bl][Bl][.8]{{\sf Angle [degree]}}
\psfrag{Rl=1}[Bl][Bl][.8]{$\tilde{R} = 1$}
\psfrag{Rl=2}[Bl][Bl][.8]{$\tilde{R} = 2$}
\psfrag{Rl=8}[Bl][Bl][.8]{$\tilde{R} = 8$}
\includegraphics[width=12.5cm,clip]{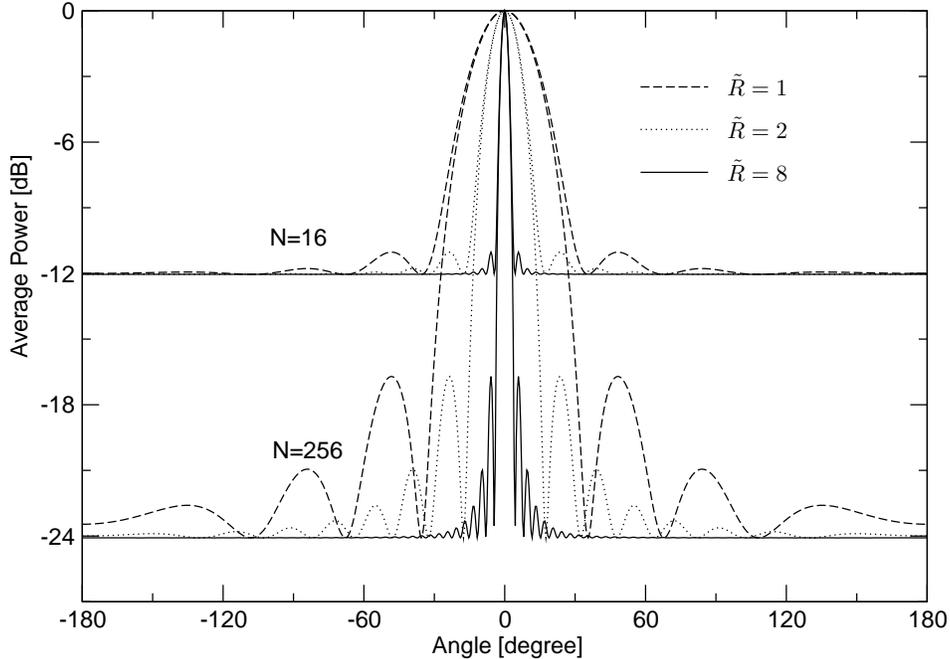}
\caption{Average beampattern with different $\tilde{R}$ and $N=16, 256$.}
\label{fig:avrage_beam}
\end{center}
\vspace{-1cm}
\end{figure}

The average beampattern of \eqref{av_beam1} is plotted in
Fig.\,\ref{fig:avrage_beam} 
for several values of $\tilde{R}$ with $N=16$ and 256.
As can be observed, the sidelobe approaches $1/N$ as the 
beam angle moves away from the target direction.

To gain further insight, consider the asymptotic expansion of
the Bessel function $J_1(x)$ for $x \gg 1$ as
\begin{equation}
J_1(x)  \sim \sqrt{\frac{2}{\pi x}}\cos\left(x - \frac{3}{4}\pi \right) .
\label{bessel_asympto}
\end{equation}
We then have
\begin{equation}
\left|2 \cdot \frac{J_1(x)}{x} \right|^2 \sim
\frac{8}{\pi x^3} 
\cos^2 \left(x - \frac{3}{4}\pi \right), 
\label{bssel_asympto2}
\end{equation}
and \eqref{av_beam1} becomes, for $\alpha(\phi) = 4\pi\tilde{R}\sin
\left(\frac{\phi}{2}\right) \gg 1$,
\begin{align}
{P}_{\text{av}}( \phi )
 & \sim 
\frac{1}{N} + 
\left( 1 - \frac{1}{N} \right)
\frac{8}{\pi \alpha(\phi)^3}
\cos^2 \left( \alpha(\phi) - \frac{3}{4}\pi \right) .
\end{align}

The $n$th peak of the average sidelobe will appear
around $\alpha(\phi_n) \approx \left(n
-\frac{1}{4} \right) \pi$, $n=1,2,\ldots$, 
and its corresponding value becomes
\begin{equation}
{P}_{\text{av}}(  \phi^{\text{peak}}_n ) \sim
\frac{1}{N} + \left( 1 - \frac{1}{N} \right)
\frac{1}{\pi} \left[
\frac{2}{\pi \left(n - \frac{1}{4}\right)}
\right]^3 ,
\label{peak_nth}
\end{equation}
which does not depend on $\tilde{R}$. 
The $n$th peak and $n$th zero positions 
(in the sense of the second term
in \eqref{av_beam1}) can then be expressed asymptotically as
\begin{align}
 \phi^{\text{peak}}_n 
& \sim 2 \arcsin\left( \frac{n - \frac{1}{4}}{4\tilde{R}}
			 \right)
\label{peak_position}
\\ 
 \phi^{\text{zero}}_n 
& \sim 2 \arcsin\left( \frac{n + \frac{1}{4}}{4\tilde{R}}
			 \right) .
\label{zero_position}
\end{align}
Since the peak sidelobe value does not depend on
$\tilde{R}$ and is less sensitive to the value of $N$, 
it is apparent that the only way one can avoid high peaks
in the sidelobe region is to increase
$\tilde{R}$ such that most of the major peaks are relatively
concentrated around the mainlobe. This phenomenon will be further 
examined in the following subsections.

\subsection{3\,dB Beamwidth of the Average Beampattern}

One of the important figures of merit in directional
antenna design is the 3\,dB beamwidth. In the deterministic
antenna, the 3\,dB beamwidth
is the threshold angle at which the power of the 
beampattern drops 3\,dB below that in the target direction 
$\phi_0$. In our scenario, the 3\,dB beamwidth itself is a random
variable and it is not easy to characterize analytically. 
Thus, as an alternative measure, we may define
the 3dB beamwidth of the {\em average} beampattern
denoted by $\phi^{3\text{dB}}_{\text{av}}$
as the angle $\phi$ that satisfies
\begin{equation}
P_{\text{av}}(\phi^{3\text{dB}}_{\text{av}}) = \frac{1}{2} .
\end{equation}
In the limit as $N \to \infty$, 
one may obtain 
\begin{equation}
\phi^{3\text{dB}}_{\text{av}} 
= 2 \arcsin \left( \frac{0.1286}{\tilde{R}}
\right) ,
\label{3dBbeamwidth}
\end{equation}
by numerically solving \eqref{av_beam1}.
For $\tilde{R} \gg 1$, 
\eqref{3dBbeamwidth} can be approximated
as $\phi^{3\text{dB}}_{\text{av}} \approx 0.26 / \tilde{R}$.
Therefore, the beamwidth is asymptotically independent of
$N$ and is mainly determined by the inverse of the
disk radius of the cluster. Consequently, sparsely
distributed sensors form a narrow beam on average provided
that the cluster radius is sufficiently large.

This sharp mainbeam property may be desirable, but
if the far-field destination node has mobility,
it should be designed carefully; the calibration should
take place before the mobile node moves out of the
mainbeam, but the mainbeam width is inversely
proportional to the normalized radius $\tilde{R}$
as observed in Fig.\,\ref{fig:avrage_beam}.
Therefore, calibration should be performed 
more frequently if the destination node moves rapidly
or when $\tilde{R}$ is increased.

\subsection{3\,dB Sidelobe Region}
\label{sec:3dBsidelobe_region}

\begin{figure}[t]
\begin{center}
\psfrag{1}[r][r][1.0]{{\sf 1}}
\psfrag{1/2}[r][r][1.0]{$\frac{\text{\sf 1}}{\text{\sf 2}}$}
\psfrag{2/N}[r][r][1.0]{$\frac{\text{\sf 2}}{\text{\sf N}}$}
\psfrag{1st peak}[B][B][.8]{{\sf 1st peak}}
\psfrag{n0th peak}[B][B][.8]{{\sf $n_0$th peak}}
\psfrag{1st zero}[B][B][.8]{{\sf 1st zero}}
\psfrag{n0th zero}[B][B][.8]{{\sf $n_0$th zero}}
\psfrag{3dB beamwidth}[T][T][.8]{{\sf 3dB beamwidth}}
\psfrag{mainbeam}[B][B][.8]{{\sf mainbeam}}
\psfrag{3dB sidelobe region}[B][B][.8]{{\sf 3dB sidelobe region}}
\includegraphics[width=12.5cm,clip]{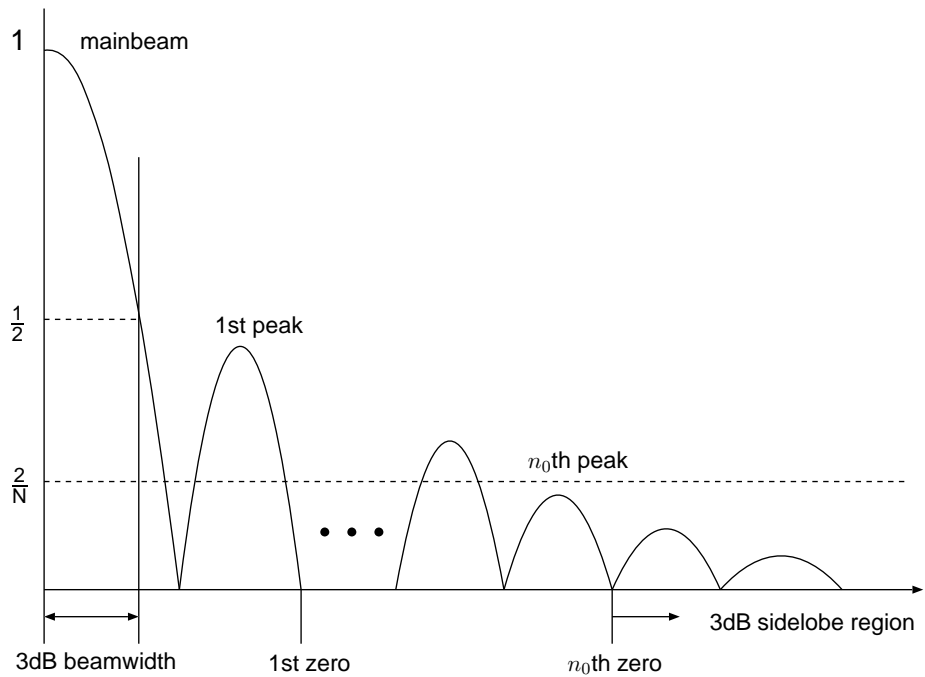}
\caption{Definitions of 3\,dB beamwidth and 3\,dB sidelobe region.}
\label{fig:3dBdefs}
\end{center}
\vspace{-1cm}
\end{figure}

Similar to the 3\,dB beamwidth concept, 
it may also be convenient for our subsequent analysis
to define the region within
which the average of the sidelobe beampattern
falls below some threshold level. 
As we have seen, for large $\tilde{R}$,
the sidelobe of the average beampattern 
is given by $1/N$ asymptotically. Therefore, we shall define the
{\em 3\,dB sidelobe region} as the region
in which neither neighboring sidelobe peak in the average beampattern
exceeds 3\,dB above $1/N$.
Let $n_0$ denote the minimum index of the 
peak position such that the corresponding peak value
satisfies this 3\,dB condition. Specifically,
from \eqref{peak_nth}, 
$n_0$ is the minimum integer $n$ that satisfies
\begin{align}
\frac{{P}_{\text{av}}(  \phi^{\text{peak}}_n ) }{1/N} \sim
1  + \left( N - 1  \right)
\frac{1}{\pi} \left[
\frac{2}{\pi \left(n - \frac{1}{4}\right)}
\right]^3 & \leq 2 ,
\label{3dB_region_def}
\end{align}
and it can be bounded by
\begin{align}
n_0 & \geq \frac{1}{4}
+ \frac{2}{\pi} 
\left(
\frac{N-1}{\pi}
\right)^{\frac{1}{3}} .
\end{align}
Let $\phi^{\text{zero}}_{n_0} > 0$ denote the angle corresponding
to the zero point next to the $n_0$th peak sidelobe which can 
be obtained from
\eqref{zero_position} with $n=n_0$.
Consequently, in this paper, the 3\,dB sidelobe region
is defined as
\begin{align}
{\cal S}_{3\text{dB}} & \triangleq
\left\{
\phi  \, \left| \,
\pi \geq  |\phi| \geq \phi^{\text{zero}}_{n_0} 
\right.
\right\} .
\end{align}
Fig.\,\ref{fig:3dBdefs} illustrates the definitions
of the 3\,dB sidelobe region together with that of the 3\,dB beamwidth.

As will be shown in Section \ref{sec:zero_mean},
the idea behind the introduction of 3\,dB sidelobes is 
that in this region one may assume that the mean value
of the random array factor of \eqref{far_field_F3}
sampled at $\phi \in {\cal S}_{3\text{dB}}$ 
becomes a random variable with approximately zero mean.
Thus we may reasonably assume that the array factor has zero mean
in this region, and this assumption significantly simplifies
the analysis of the statistical distribution in the 
following sections.

\begin{figure}[t]
\begin{center}
\psfrag{R/l}[Bl][Bl][.8]{$\tilde{R}$}
\psfrag{Minimum Angle [degree]}[Bl][Bl][.8]{{\sf Threshold Angle [degree]}}
\includegraphics[width=12.5cm,clip]{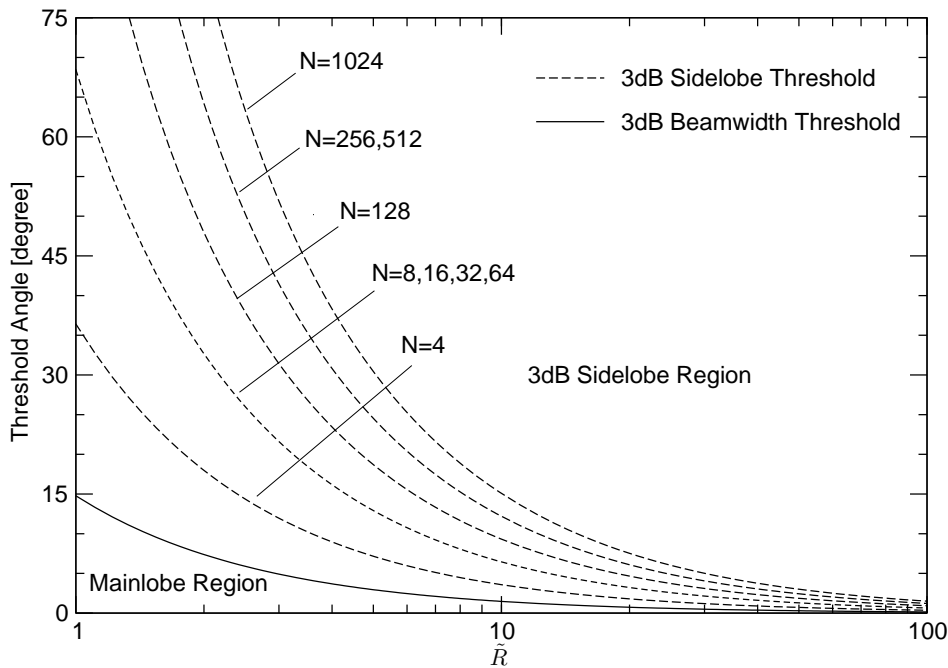}
\caption{Threshold of 3\,dB beamwidth and 3\,dB
sidelobe region with respect to $\tilde{R}$
and number of nodes $N$.}
\label{fig:three_side}
\end{center}
\vspace{-1cm}
\end{figure}

Fig.\,\ref{fig:three_side} 
shows the threshold angle above which the 3\,dB sidelobe region
begins. The asymptotic 3\,dB beamwidth 
\eqref{3dBbeamwidth} is also shown for reference. 
As can be observed, whereas the 3\,dB beamwidth 
is less sensitive to 
the number of nodes $N$,
the 3\,dB sidelobe region will be considerably reduced
as $N$ increases. This means that
as $N$ increases 
the dominant non-negligible sidelobe peak 
may occur with high probability
unless $\tilde{R}$ is also increased. This trade-off
will be clarified by the study of directivity 
in the following subsection.

\subsection{Average Directivity}

The directivity or directional gain is the parameter that
characterizes how much radiated energy is concentrated
in the desired direction, relative to a single 
isotropic antenna. 
Specifically, it may be defined as
\begin{equation}
D \triangleq \frac{\int_{-\pi}^{\pi} P(0) d \phi}{
\int_{-\pi}^{\pi} P(\phi) d \phi} 
= \frac{2 \pi }{
\int_{-\pi}^{\pi} P(\phi) d \phi} ,
\end{equation}
where $P(\phi)$ is the radiated power density
in the direction of $\phi$.
In the scenario of this paper, since $P(\phi)$ depends
on the particular realization of ${\boldsymbol z}$,
the corresponding gain may be expressed, by substituting
$P(\phi | {\boldsymbol z})$ of \eqref{beampattern1} into
the above, as
\begin{equation}
D ({\boldsymbol z})
 = \left[
{ \frac{1}{N} +  
\frac{1}{N^2} \sum_{k=1}^N 
\sum_{\stackrel{ \scriptstyle l=1}{l\neq k}}^{N} 
J_0\left( 4 \pi \tilde{R} (z_k - z_l) \right)}
\right]^{-1} .
\label{directivity}
\end{equation}
The mean value of \eqref{directivity} is given by
\begin{equation}
D_{\text{av}} \triangleq E_{\boldsymbol z}
\left\{
D(\boldsymbol z)
\right\}  .
\label{lower_Dav}
\end{equation}
Unfortunately, direct calculation of 
\eqref{lower_Dav} does not result in a closed-form or
insightful expression. Thus,
we shall consider the following as an alternative measure:
\begin{equation}
\tilde{D}_{\text{av}} \triangleq 
\frac{2\pi}
{ \int_{-\pi}^{\pi} E_{\boldsymbol z}
\left\{
P(  \phi | {\boldsymbol z})
\right\}
d \phi}
= \frac{2 \pi}
{ \int_{-\pi}^{\pi} P_{\text{av}}(  \phi) d \phi} .
\label{Dav1}
\end{equation}
Note that by Jensen's inequality, we have 
\begin{equation}
\tilde{D}_{\text{av}} \leq D_{\text{av}},
\label{jensen_relation}
\end{equation}
which means that $\tilde{D}_{\text{av}}$ in \eqref{Dav1}
is a lower bound on $D_{\text{av}}$. However, since 
by the law of large numbers 
the denominator of $D(\boldsymbol z)$ may
approach its average value with high probability as $N$ increases, 
the above bound is expected to become tight as $N$ increases.
This will be verified in the numerical results below.
Substituting \eqref{av_beam1} into \eqref{Dav1},
we obtain
\begin{equation}
\tilde{D}_{\text{av}} = \frac{N}
{1 + (N-1) {\,}_2 F_3 
\left( \frac{1}{2}, \frac{3}{2} \,;\, 1, 2, 3 \,;\,
- ( 4 \pi \tilde{R} )^2
\right) } , 
\label{Dav2}
\end{equation}
where ${\,}_2 F_3 
\left( \frac{1}{2}, \frac{3}{2} \,;\, 1, 2, 3 \,;\,
- x^2 \right)$ is a generalized hypergeometric function
which monotonically decreases with increasing $x$ 
and converges to 0 as $x \to \infty$.
Therefore, unlike well-designed deterministic linear arrays,
the gain of a given realization is very likely to be less than $N$, 
and the limit $N$ can be approached only by increasing $\tilde{R}$.
This agrees with the previous observation that the average
mainbeam becomes narrow as $\tilde{R}$ increases
and thus improves the directivity.

It should be noted that although \eqref{Dav2} 
has a simple form and offers some insight into the
asymptotic behavior of directivity,
the calculation of the generalized hypergeometric function involved
in \eqref{Dav2} becomes numerically unstable as 
$\tilde{R}$ increases, and it is much easier to 
numerically integrate the denominator of \eqref{Dav1} directly.

\begin{figure}[t]
\begin{center}
\psfrag{Relative Directivity D/N}[Bl][Bl][.8]{{\sf Normalized
 Directivity} $\tilde{D}_\text{av} / N$}
\psfrag{Normalized Radius R/lambda}[Bl][Bl][.8]{{\sf Normalized Radius }$R/\lambda$}
\includegraphics[width=12.5cm,clip]{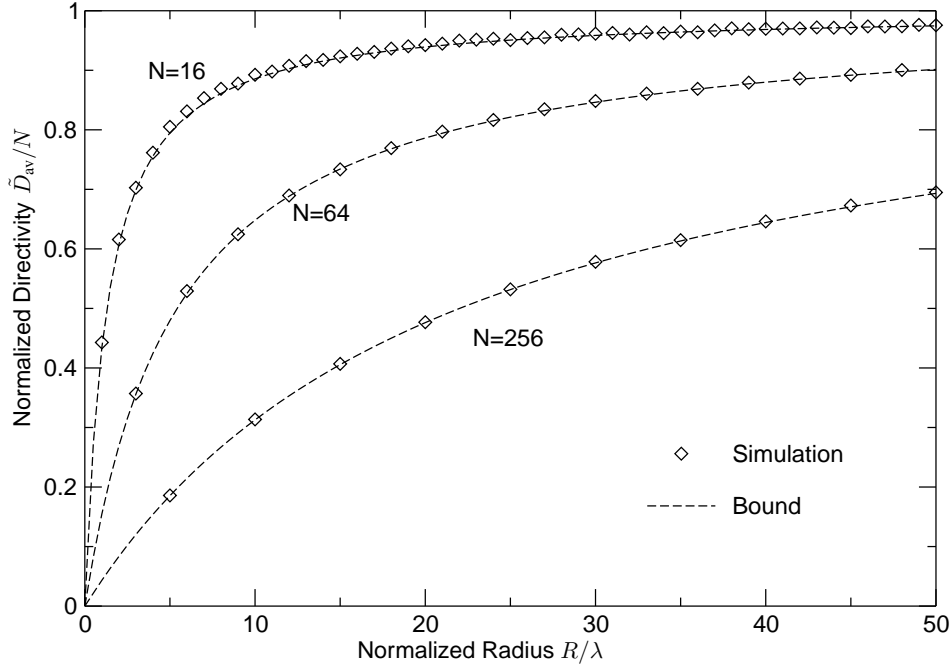}
\caption{The relationship between directivity $\tilde{D}_{\text{av}}/N$ 
and normalized radius $\tilde{R}$.}
\label{fig:directivity}
\end{center}
\vspace{-1cm}
\end{figure}

Fig.\,\ref{fig:directivity} shows the relationship between the
normalized directivity bound $\tilde{D}_{\text{av}}/N$ 
and $\tilde{R}$. Also shown
in the figure as diamond-shaped points
are the corresponding exact average directivities ${D}_{\text{av}}/N$
obtained by the simulation of 1000 realizations.
As can be observed, the bound is very tight compared to the exact performance.
Thus, it follows that in order to achieve high normalized directivity
(i.e. directivity close to $N$) with $N$ nodes,
the distribution of the nodes should be as sparse as possible.
In fact, we have the following theorem:

\begin{theorem}[Normalized Directivity Lower Bound]
\label{theorem_directivity}
For large $\tilde{R}$ and $N$, $D_{\text{av}}/N$ is lower bounded
by
\begin{equation}
\frac{D_{\text{av}}}{N} \geq \frac{1}{1 + \mu \frac{N}{\tilde{R}}},
\label{theorem1_eq}
\end{equation} 
where $\mu$ is a positive constant independent of $N$ and $\tilde{R}$
($\mu \approx 0.09332$).
\end{theorem}

\begin{proof}
See Appendix \ref{app:directive}.
\end{proof}

\begin{figure}[t]
\begin{center}
\psfrag{Normalized Directivity}[Bl][Bl][.8]{{\sf Normalized Directivity
 }$\tilde{D}_{\text{av}} / N$}
\psfrag{N=16}[Bl][Bl][.8]{{\sf Exact (N=16)}}
\psfrag{N=256}[Bl][Bl][.8]{{\sf Exact (N=256)}}
\psfrag{Theorem 1}[Bl][Bl][.8]{{\sf Lower Bound (Theorem 1)}}
\psfrag{Node Density}[Bl][Bl][.8]{{\sf Node Density }$N / \tilde{R}$}
\includegraphics[width=12.5cm,clip]{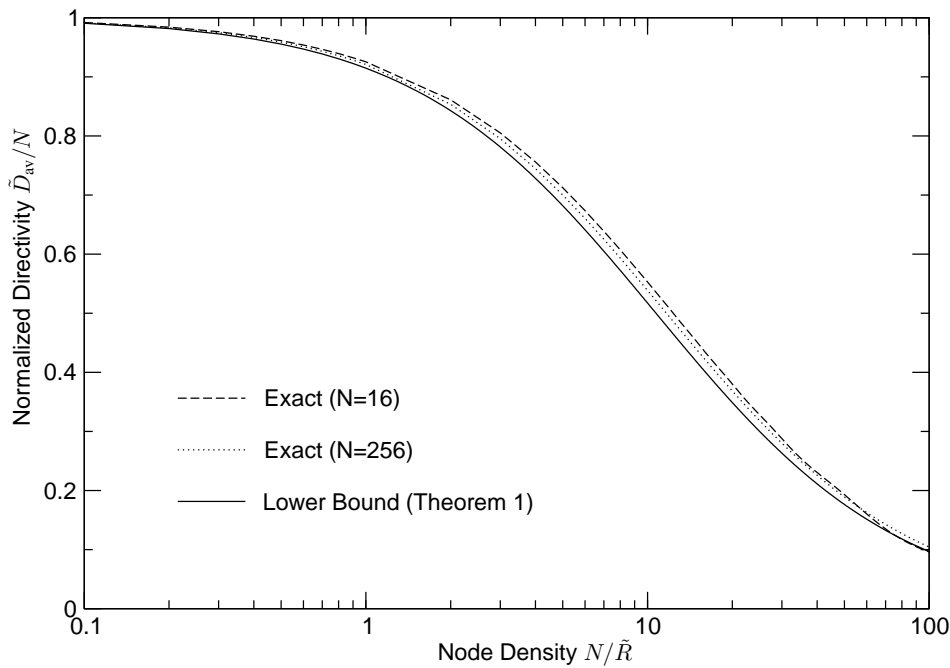}
\caption{The relationship between directivity $\tilde{D}_{\text{av}}/N$ and 
 node density $N/\tilde{R}$.}
\label{fig:directivity_density}
\end{center}
\vspace{-1cm}
\end{figure}

Note that the factor $N/\tilde{R}$ 
which appears in \eqref{theorem1_eq}
can be seen as a 
{\em one-dimensional node density}. 
To verify the above theorem, Fig.\,\ref{fig:directivity_density} 
shows the relationship between $\tilde{D}_{\text{av}}/N$ and the 
node density $N/\tilde{R}$,
as well as the lower bound in \eqref{theorem1_eq}.

The above theorem indicates that
there is a simple relationship between directivity and node density.
It can be seen that the node density
almost uniquely determines the normalized directivity $D_{\text{av}}/N$.
It is important to note that in order to achieve a certain 
normalized directivity with large numbers of
nodes $N$, the node density
should be maintained to the desired value
by spreading the nodes as sparsely as possible. Alternatively, 
if the normalized region $\tilde{R}$ is fixed, it is not efficient 
in terms of normalized directivity to increase the number of sensor nodes.

The above theorem also indicates that if the sensor nodes are 
uniformly distributed and 
if we are to choose $N$ nodes out of them, 
in terms of normalized directivity it may be
better to choose them as sparsely as possible.

\section{Distribution of Far-Field Beampattern of Collaborative
Beamforming with Perfect Phase Information}
\label{sec:distribution_beam}

\begin{figure}[t]
\begin{center}
\includegraphics[width=12.5cm,clip]{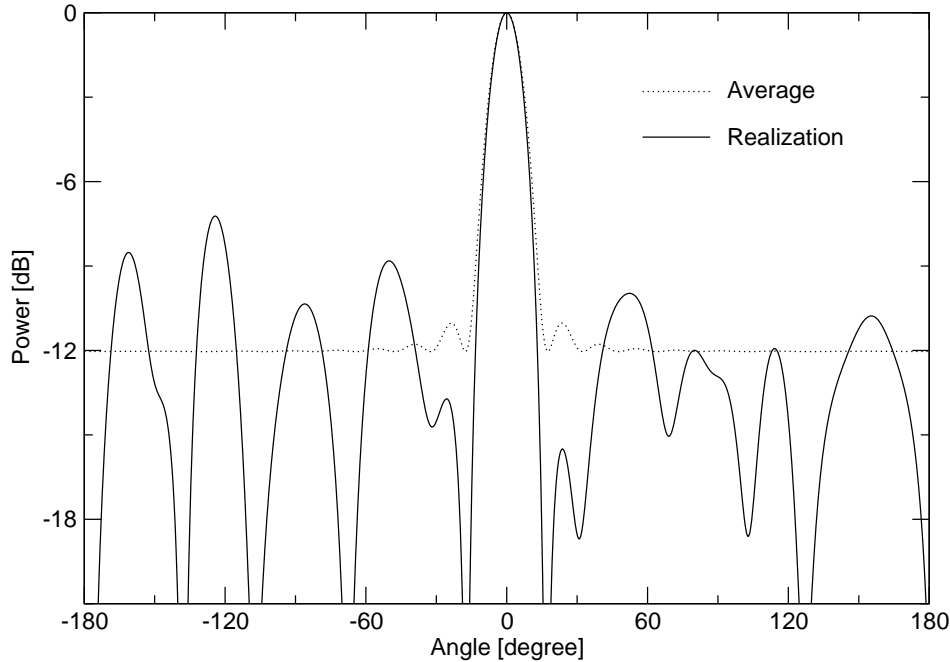}
\caption{Average and realization of beampattern with $\tilde{R}=2$ and $N=16$.}
\label{fig:av_vs_instant_beam}
\end{center}
\vspace{-1cm}
\end{figure}

In the previous section, we have seen that random arrays
have nice average beampatterns with low sidelobes.
However,  the average behavior does not necessarily
approximate a beampattern of any given realization unless $N\to\infty$.
In fact, even though 
the average beampattern has a sharp mainbeam and 
sidelobes always close to $1/N$, 
there is a large dynamic range of sidelobes
among randomly generated beampatterns. 
As an example, the average beampattern and one particular realization
of randomly generated sensor locations is shown in
Fig.\,\ref{fig:av_vs_instant_beam}. The mainbeam 
of the realization 
closely matches the average, but sidelobes may fluctuate
with a large dynamic range and easily exceed the average level.

Therefore, in practice, the statistical {\em distribution} 
of beampatterns and sidelobes in particular, is of interest.
By approximating the beampattern sidelobes as 
a complex Gaussian process, 
Lo \cite{Lo:1967} has derived 
the distribution of the 
beampattern in the case of linear random arrays.

In the following, we first derive a numerical method
that allows calculation of the exact distribution of the beampattern
without applying Gaussian approximations.
We then derive a convenient asymptotic form of the sidelobe
distribution using a Gaussian approximation
similar to \cite{Lo:1967}, and evaluate its validity in our framework.

\subsection{Exact Evaluation of Distribution}
\label{sec:exact_dist}

Since the array factor is a sum of i.i.d. random variables, its
distribution can be 
computed numerically by the characteristic function
method. To this end, 
from \eqref{far_field_F4} let
\begin{align}
\tilde{F}(\phi | {\boldsymbol z} )
&  =
\frac{1}{N} \sum_{k=1}^{N} \left( \tilde{x}_k
- j  \tilde{y}_k \right)
\triangleq \frac{1}{N} \left( \tilde{X} - j \tilde{Y} \right) ,
\label{Fsum}
\end{align}
where 
\begin{align}
\tilde{x}_k & \triangleq
\cos\left(
 z_k 
\alpha(\phi)
\right), 
\quad
\tilde{y}_k  \triangleq
\sin\left(
 z_k
\alpha(\phi)
\right)
\end{align}
and $\alpha(\phi)$ is defined in \eqref{alpha_def}.
The joint characteristic function of $\tilde{x}_k$ and $\tilde{y}_k$
is written as
\begin{align}
\varPhi_{\tilde{x}_k, \tilde{y}_k} (\omega, \nu) & = 
E_{\tilde{x}_k, \tilde{y}_k} 
\left\{
e^{j \left(
\omega \tilde{x}_k + \nu \tilde{y}_k \right)}
\right\}
 = E_{z_k} \left\{
e^{j 
\left[
\omega \cos\left(
 z_k \alpha(\phi)
\right)
+ \nu 
\sin\left(
 z_k  \alpha(\phi)
\right)
\right]}
\right\} .
\label{joint_CHF1}
\end{align}
For a given pair of $\omega$ and $\nu$, the above expectation
is a single integral of a well-behaved function
and can be calculated numerically.

Since $\tilde{F}(\phi | {\boldsymbol z} )$ is a sum
of $N$ i.i.d. complex random variables, the joint 
probability density of $\tilde{X}$ and $\tilde{Y}$ in \eqref{Fsum}
is given by 
\begin{equation}
f_{\tilde{X},\tilde{Y}}( x, y )  = \left( \frac{1}{2\pi}\right)^2
\int_{-\infty}^{\infty} \int_{-\infty}^{\infty}
\left[
\varPhi_{\tilde{x}_k, \tilde{y}_k}
(\omega, \nu) 
\right]^N 
e^{-j(\omega x + \nu y)}
\,d\omega \, d\nu .
\end{equation}
The above integral can be computed efficiently 
using the two-dimensional Fast Fourier Transform (FFT).
Finally, the complementary cumulative distribution function
(CCDF) of the beampattern, i.e., the
probability that the instantaneous power of a given realization
in the direction $\phi$ exceeds a threshold power, $P_0$, is given by
\begin{align}
\Pr \left[ P( \phi) > P_0
\right] & = \Pr \left[ \frac{\tilde{X}^2 + \tilde{Y}^2}{N^2} > P_0 
\right] 
= \int \!\!\!  \int_{x^2 + y^2 > N^2 P_0 } f_{\tilde{X}, \tilde{Y}} 
(x,y) \,dx\, dy .
\label{exact_dist_num}
\end{align}

\subsection{Gaussian Approximation of Distribution}
The exact evaluation of the CCDF outlined above is computationally demanding,
especially if the desired numerical precision is high.
Considering that the array factor consists of a sum of $N$ statistically
independent random variables, as $N$ increases, 
by the central limit theorem we may expect that the
array factor with any given direction, except at the deterministic
angle $\phi=0$, approaches a complex Gaussian distribution. This approximation
may typically
result in a simpler distribution formula. To this end,
we write \eqref{Fsum} as
\begin{align}
\tilde{F}(\phi | {\boldsymbol z} )
&  =
\frac{1}{\sqrt{N}}  \left(
{X} - j {Y} \right)
\label{Frev}
\end{align}
where
\begin{align}
{X} & \triangleq
\frac{1}{\sqrt{N}}
\sum_{k=1}^{N}
\cos\left(
 z_k
\alpha(\phi)
\right), 
\quad
{Y}  \triangleq
\frac{1}{\sqrt{N}}
\sum_{k=1}^{N}
\sin\left(
 z_k \alpha(\phi)
\right) .
\label{XYgauss}
\end{align}
Since the $z_k$'s are i.i.d. random variables,
as $N$ increases the distribution of ${X}$ and ${Y}$
at the direction $\pi \geq |\phi| > 0$
may approach that of a Gaussian random variable with
\begin{align}
E\left\{
X
\right\}
 & =  
 2 \frac{J_1\left(
\alpha (\phi) \right)}
{\alpha (\phi)} \sqrt{N} \triangleq m_x 
\label{gauss_mean} \\
\text{Var}
\left(
X
\right)
 & =
\frac{1}{2}\left( 1 + 
 \frac{J_1\left(
2 \alpha (\phi)  \right)}
{\alpha (\phi)} 
\right) -
\left[
2  \frac{J_1\left(
 \alpha  (\phi) \right)}
{\alpha (\phi)}
\right]^2
\triangleq \sigma^2_x
\label{gauss_var_x}  \\
E
\left\{
Y
\right\}
 & =  0 
\label{gauss_mean_y} \\
\text{Var}
\left(
Y
\right)
 & =
\frac{1}{2}\left( 1 -
 \frac{J_1\left(
2 \alpha (\phi)  \right)}
{\alpha (\phi)} 
\right) \triangleq \sigma^2_y  .
\label{gauss_var_y}
\end{align}
Note that 
$E\left\{ {X} \, {Y}
\right\} =0$, i.e., 
${X}$ and ${Y}$ 
are orthogonal and thus statistically uncorrelated.
The joint pdf of ${X}$ and ${Y}$ is then given by
\begin{equation}
f_{{X},{Y}}(x, y) = 
\frac{1}{2 \pi \sigma_x \sigma_y } \exp
\left(
-\frac{| x - m_x |^2 
}{2\sigma^2_x}
-\frac{ y^2}{2 \sigma^2_y}
\right) .
\label{exact_pdf_gauss}
\end{equation}
The CCDF of ${P}_0$ can be expressed as
\begin{align}
\Pr \left[ P( \phi) > {P}_0
\right] 
& = \Pr \left[ 
 \frac{{X}^2 + {Y}^2}{N} > {P}_0
\right] 
= \Pr \left[ 
 \sqrt{ {{X}^2 + {Y}^2}}
> \sqrt{N {P}_0}
\right] 
\nonumber \\
& = \int_{\sqrt{N{P}_0}}^{\infty}  \int_{-\pi}^{\pi}
\frac{r}{2 \pi \sigma_x \sigma_y } \exp
\left(
-\frac{| r \cos\omega - m_x |^2 
}{2\sigma^2_x}
-\frac{ r^2 \sin^2 \omega }{2 \sigma^2_y}
\right)
d\omega \, dr
\nonumber \\
& =  \int_{-\pi}^{\pi}
\frac{1}{4 \pi \sigma_x \sigma_y U^2_{\omega}}
e^{V^2_{\omega} - \frac{m_x^2}{2 \sigma_x^2} }
\left[
\sqrt{\pi} V_{\omega} \text{ erfc}\left( W_{\omega} - V_{\omega} \right)
+ e^{-(W_{\omega} -V_{\omega} )^2}
\right]
d\omega ,
\label{exact_CCDF_gauss}
\end{align}
where
\begin{align}
U_{\omega} & \triangleq \sqrt{ \frac{\cos^2 \omega }{2 \sigma^2_x}
+  \frac{\sin^2 \omega }{2 \sigma^2_y}}, \quad
V_{\omega}  \triangleq \frac{m_x \cos \omega}{2 \sigma^2_x {U_{\omega}}}
, \quad
W_{\omega}  \triangleq \sqrt{N {P}_0} U_{\omega} .
\end{align}

For $\alpha(\phi) \gg 1$, the terms $J_1(2 \alpha (\phi)) / \alpha (\phi)$ and
$\left|J_1(\alpha (\phi)) / \alpha (\phi) \right|^2$  
in the variance expressions \eqref{gauss_var_x} and \eqref{gauss_var_y}  
rapidly decrease and their contribution to the resulting
variances becomes minor.
Therefore, it is very likely that both variances are approximately
equal in the sidelobe region. When this is the case, i.e., if
$\sigma^2_x \approx \sigma^2_y \approx 1/2$,
the distribution of the complex envelope becomes 
a Nakagami-Rice distribution. Consequently, the resulting
integral can be expressed in terms of the first-order Marcum-Q function
\begin{equation}
\Pr \left[ P( \phi) > {P}_0
\right] 
 =  Q \left(
\frac{m_x}{\sigma_x}, \frac{\sqrt{N {P}_0}}{\sigma_x}
\right) 
 =  Q \left(
\sqrt{2} {m_x},  \sqrt{2 N {P}_0}
\right) .
\label{marcum_formula}
\end{equation}

Furthermore, if the mean $E\left\{X\right\}$ is zero,
the envelope follows a Rayleigh distribution and we simply have
\begin{align}
\Pr \left[  P(\phi) > {P}_0
\right] 
& =  e^{- N {P}_0} .
\label{full_gaussian}
\end{align}

\subsection{Mean Value of Array Factor within 3\,dB Sidelobe Region}
\label{sec:zero_mean}

As we have seen, if the mean value of the array factor can be assumed to be
zero, the distribution can be significantly simplified and thus analysis
becomes readily tractable. 
From \eqref{gauss_mean} it is apparent that under the constant variance
constraint the mean value increases as $N$ increases.
Therefore, when $N$ is large, the zero mean assumption may not
be guaranteed in general.
In Section \ref{sec:3dBsidelobe_region},
we have introduced the 3\,dB
sidelobe region, and in the following we derive properties of the mean
value of the array factor in this region.

From the definition of \eqref{3dB_region_def}, the sidelobe in 
the 3\,dB region satisfies
\begin{align}
& N {P}_{\text{av}} (\phi)  \leq 2 .
\end{align}
It follows that
\begin{align}
& \text{Var} 
\left( X \right)
 +  \text{Var} 
\left( Y \right)
 + 
\left| E \left\{
X
\right\}
\right|^2  \leq 2 .
\end{align}
From \eqref{gauss_mean}, \eqref{gauss_var_x}, and \eqref{gauss_var_y},
we have
\begin{align}
& 1  - \frac{1}{N} \left| E
\left\{ X \right\}
 \right|^2 
+ \left| E \left\{ X \right\}
\right|^2  \leq 2,
\end{align}
and thus we get
\begin{align}
&  \left| E
\left\{ X \right\}
 \right|^2  \leq  \frac{1}{1 - \frac{1}{N}} .
\end{align}
Therefore, in the 3\,dB sidelobe region,
the square of the mean is bounded by unity
in the large-$N$ asymptote and thus
the mean does not grow unbounded with the number of nodes
$N$.

\subsection{Numerical Comparison}

\begin{figure}[t]
\begin{center}
\includegraphics[width=12.5cm,clip]{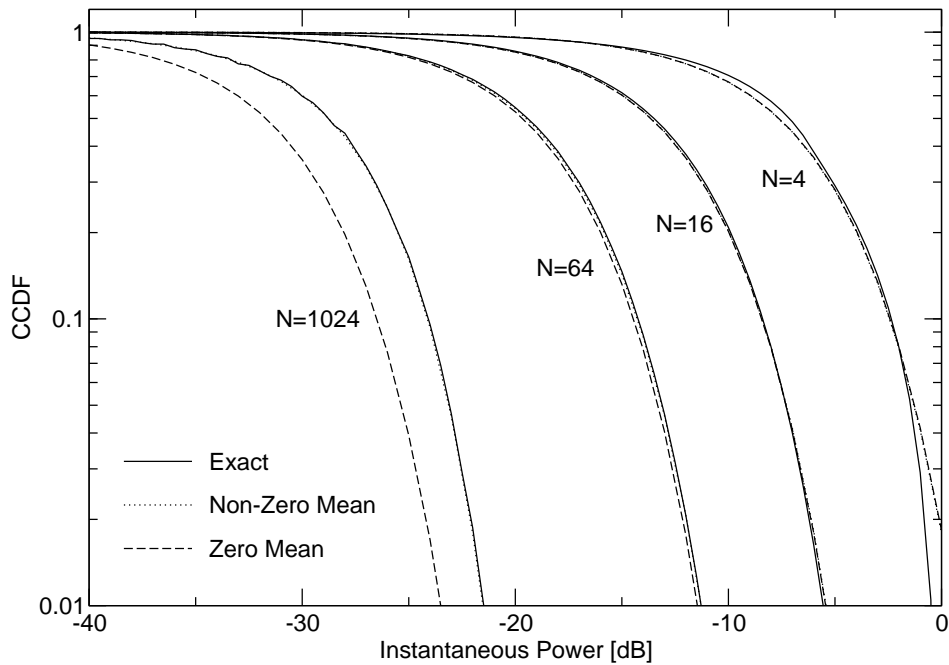}
\caption{CCDF of beampattern with $\phi=\pi/4$, and $\tilde{R}
 = 2$.}
\label{fig:distribution_r2}
\end{center}
\vspace{-1cm}
\end{figure}

In Fig.\,\ref{fig:distribution_r2}, the CCDF's computed with various 
formulae are shown with $\tilde{R}=2$
and $ \phi=\pi/4$, which corresponds to the sidelobe region. 
The exact formula of \eqref{exact_dist_num},
the equal variance Gaussian approximation of \eqref{marcum_formula},
and the zero-mean Gaussian approximation of \eqref{full_gaussian}
are shown in the figure. Note that the precise Gaussian case of
\eqref{exact_CCDF_gauss} was also calculated by numerical integration,
but it is almost identical to the Marcum-Q function approximation
in \eqref{marcum_formula} for this case
and thus is not shown.
As observed from Fig.\,\ref{fig:distribution_r2},
even the zero-mean Gaussian approximation may be valid for 
this sidelobe region, but for $N=1024$ the Gaussian approximation 
will have some noticeable discrepancy 
with the exact value.
This is due to the fact that the zero-mean 
approximation does not hold for this case. In fact, 
Fig.\,\ref{fig:three_side} indicates that 
for this value of $N$, the angle falls between the 3\,dB sidelobe 
region and the mainlobe region and thus the zero mean assumption
may not be accurate.

\begin{figure}[t]
\begin{center}
\psfrag{Exact FFT}[l][l][.8]{{\sf Exact}}
\psfrag{Numerical Gaussian}[l][l][.8]{{\sf Precise Gaussian}}
\psfrag{Marcum-Q}[l][l][.8]{{\sf Marcum-Q}}
\includegraphics[width=12.5cm,clip]{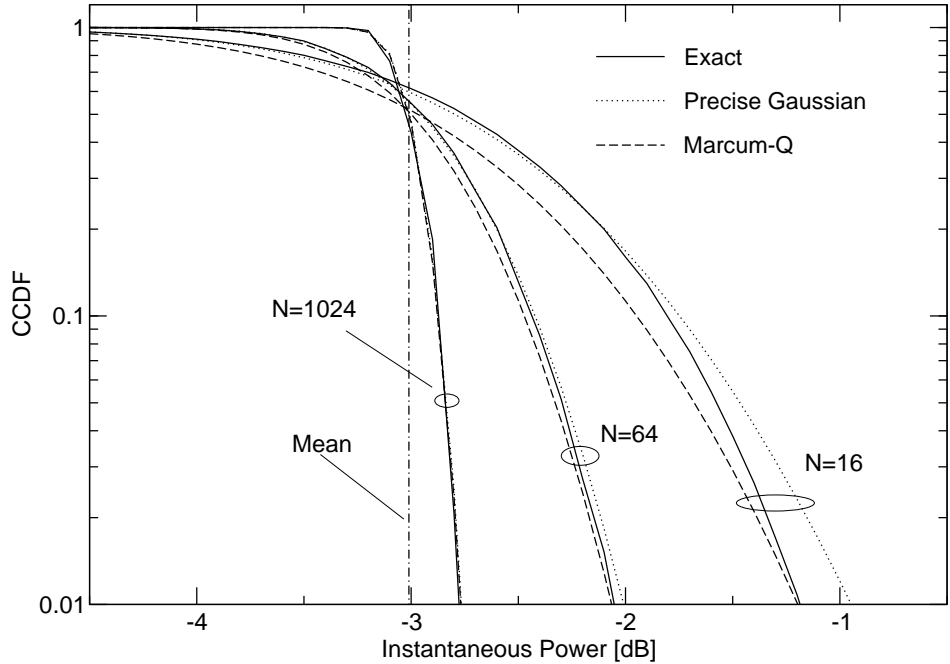}
\caption{Distribution of beampattern at
$\phi = \phi^{3\text{dB}}_{\text{av}}$ with exact,
precise Gaussian, and Marcum-Q formulas.}
\label{fig:three_main}
\end{center}
\vspace{-1cm}
\end{figure}

Fig.\,\ref{fig:three_main} shows the distribution
at 3\,dB beamwidth of the average 
beampattern defined by \eqref{3dBbeamwidth}. In this case,
the exact form \eqref{exact_dist_num}, 
the precise Gaussian \eqref{exact_CCDF_gauss}, and
the Marcum-Q approximation \eqref{marcum_formula} 
show different results even for relatively large $N$,
since at this angle the variance of the array factor 
is small and a large number of random variables
must be summed in \eqref{XYgauss} for a non-zero mean 
Gaussian random variable to adequately approximate
the sum. 
As observed, as the number of nodes increases,
the mainbeam variance becomes small and approaches 
the mean value of -3\,dB, as expected. Therefore, 
for large $N$ the mainbeam can be
made stable. This observation agrees with a similar result
for linear arrays in \cite{Lo:1967}.

\section{Distribution of the Maximum of the Sidelobes}
\label{sec:distribution_maximum}

It is well known that unlike periodic or equally-spaced
antenna arrays, many arrays with unequal spacing 
will yield no grating lobes for a large number of elements. 
This property is also preserved for random 
arrays \cite{Lo:1967,Steinberg:1976},
but in order to verify this, one may need to find
the distribution of the maximum power of the sidelobes.
In this section, we develop an approximate upper bound
on the distribution of the peak sidelobes for
random sensor networks.

In the previous section, we have seen that 
the distribution of the beampattern samples within 
the sidelobe region can be characterized by a zero mean 
Gaussian random variable
if the zero-mean condition is satisfied.
In the following, we further assume that the beampattern
is a Gaussian random {\em process}. In this case, any two samples
taken from the beampattern should be characterized by jointly
Gaussian random variables.
In the linear random array framework, the distribution of peak sidelobes
has been studied in \cite{Steinberg:1972,Agrawal:1972,Donvito:1979},
assuming the array factor is a Gaussian process.
For simplicity, only the 3\,dB sidelobe region is considered
and it will be assumed that the process is stationary with zero
mean. The extension to
the non-stationary case is studied in \cite{Donvito:1979}.

In the following, the CCDF of the maximum peak sidelobe,
which is the probability that the maximum peak sidelobe 
exceeds a given power level, 
will also be referred to as {\em outage probability} and 
denoted by $P_\text{out}$.

\subsection{Upper Bound on the Distribution of Peak Sidelobe}

Let $\nu(a)$ denote the random variable representing the 
number of upward crossings at a given level $a$ 
per interval in the 3\,dB sidelobe region ${\cal S}_{3\text{dB}}$.
As shown in Appendix \ref{app:level_crossing}, assuming that
the array factor in this region can be approximated as a
zero-mean Gaussian process, the mean of $\nu(a)$ is given by
\begin{equation}
E \left\{
\nu(a) 
\right\}
= 
4 \left( 1 -  \sin \frac{\phi^{\text{zero}}_{n_0}}{2}
 \right)
\sqrt{\pi} \tilde{R} a e^{-a^2} .
\label{mean_crossing2}
\end{equation}
Note that the above function monotonically decreases
with increasing $a$ only for $a > 1/\sqrt{2}$, and thus
is meaningful only in this region.
Finally, noticing that the outage probability
is the probability that at least one peak exceeds level $a$ and
is given by \cite{Donvito:1979}
\begin{align}
P_\text{out} = 
\Pr \left[
\nu(a) \geq 1
\right] & = \sum_{k=1}^{\infty}
\Pr[ \nu(a) = k
] 
\leq \sum_{k=1}^{\infty}
k \Pr[ \nu(a) = k
] = E
\left\{
 \nu(a)
\right\},
\end{align}
then \eqref{mean_crossing2} serves as an upper bound for the
outage probability for the maximum sidelobe peak for $a > 1/\sqrt{2}$. 
Thus, we obtain the CCDF upper bound as
\begin{align}
\Pr \left[ \max_{{\cal S}_{3\text{dB}}} {X}^2 + {Y}^2
 > P_0 \right] & \leq
4 \left(
1 -  \sin \frac{\phi^{\text{zero}}_{n_0}}{2}
\right)
\sqrt{\pi} \tilde{R} \sqrt{N P_0} e^{-N P_0} , \qquad 
\text{ for } \quad N P_0  > \frac{1}{2} .
\label{upperbound_sidelobe}
\end{align}

\subsection{Numerical Results}

\begin{figure}[t]
\begin{center}
\psfrag{R/N=0.5}[Bl][Bl][.8]{$N/\tilde{R}=2$}
\includegraphics[width=12.5cm,clip]{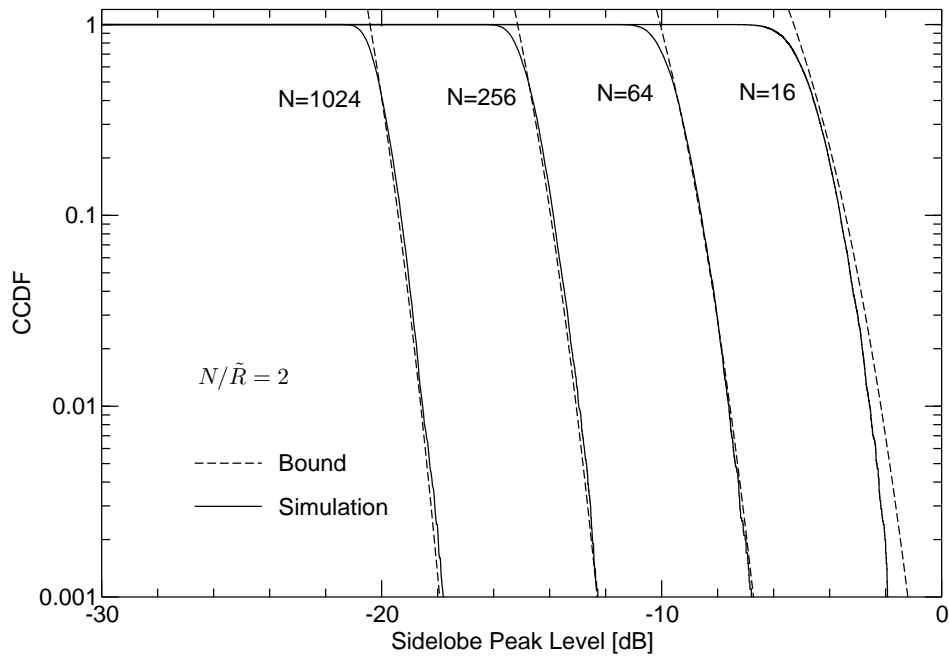}
\caption{Comparison of CCDF and upper bound of the sidelobe peaks with
the node density $N/\tilde{R}=2$.}
\label{fig:CCDF_sidelobepeak}
\end{center}
\vspace{-1cm}
\end{figure}

Fig.\,\ref{fig:CCDF_sidelobepeak} shows a comparison between
simulation results and the upper bound \eqref{upperbound_sidelobe}.
For the simulation, the outage probability is calculated based on 
10\,000 randomly generated realizations with
the node density $N/\tilde{R}=2$
and only the peaks within the 3\,dB sidelobe region are examined.
Also, in order to capture peak values accurately, the entire
3\,dB sidelobe region of $\phi$ is sampled at a rate as large as
$16 \pi \tilde{R}$.
As can be observed, the bound is in good agreement with
simulation for large $N$.

\begin{figure}[t]
\begin{center}
\psfrag{Pout=0.1\%}[Bl][Bl][.8]{$P_{\rm out} = 0.1 \%$}
\psfrag{Pout=1\%}[Bl][Bl][.8]{$P_{\rm out} = 1 \%$}
\psfrag{Pout=10\%}[Bl][Bl][.8]{$P_{\rm out} = 10 \%$}
\psfrag{Normalized Radius}[Bl][Bl][.8]{{\sf Normalized Radius}\,$\tilde{R}$}
\psfrag{Normalized Sidelobe Level [dB]}[B][B][.8]{{\sf Normalized Sidelobe Level }\,\,$\tilde{P}_0$\,\,{\sf [dB]}}
\includegraphics[width=12.5cm,clip]{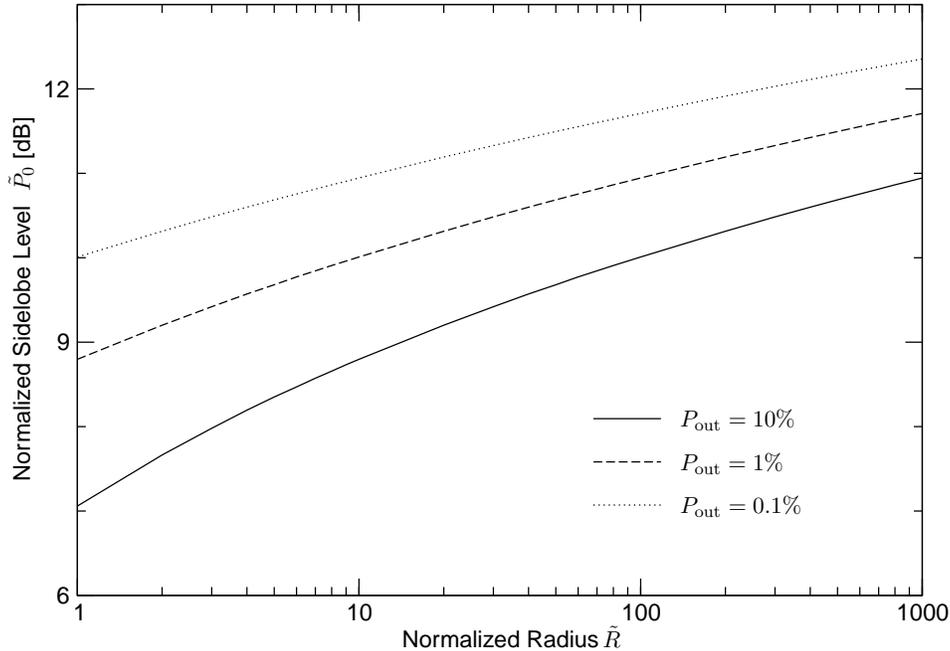}
\caption{Bound on sidelobe maximum with a given outage $P_{\text{out}}$.}
\label{fig:Outage_sidelobepeak}
\end{center}
\vspace{-1cm}
\end{figure}

Let $\tilde{P}_0 = N P_0$ denote the threshold of
the maximum peak value ($P_0$) normalized by
the average sidelobe level ($1/N$).
Since from \eqref{zero_position}
$\phi^{\text{zero}}_{n_0}$ approaches zero
as $\tilde{R}$ increases,
\eqref{upperbound_sidelobe} reduces to
\begin{align}
P_\text{out}
& \leq
4 \sqrt{\pi} \tilde{R} \sqrt{\tilde{P_0}} e^{-\tilde{P_0}},
\qquad  \tilde{P}_0 > 1/2 .
\end{align}
The above inequality illuminates the relationship
between the outage probability and $\tilde{R}$ 
(assuming that $\phi^{\text{zero}}_{n_0}$ is negligibly small).
Fig.\,\ref{fig:Outage_sidelobepeak}
shows the maximum possible value of $\tilde{P}_0$ for a given
outage probability and $\tilde{R}$. As can be observed,
the maximum sidelobe may grow as $\tilde{R}$ increases,
but the amount is below 12\,dB for many cases
of interest.
Consequently, 
the maximum sidelobe level (in the 3\,dB region) 
of randomly generated arrays may be written as
$P_0 = \tilde{P}_{0} / N$, where the
required margin $\tilde{P}_{0}$
depends on the parameters $\tilde{R}$ and $P_{\text{out}}$,
but not on $N$. Thus, increasing $N$ always results in 
a reduction of maximum sidelobe level.

\section{Performance of Distributed Beamforming with Imperfect Phase}
\label{sec:imperfect}

So far, we have evaluated the beampattern assuming perfect knowledge of
the initial phase for each node. 
In this section, we analyze the effect of the phase ambiguities
in the  closed-loop scenario
as well as location estimation errors 
in the open-loop scenario. For each of
the two scenarios, we derive the average 
beampattern and calculate the amount of mainbeam degradation. 
 
\subsection{Closed-loop Case}

In the closed-loop case, the effects of imperfect phase
may be easily derived, following the approach developed
by Steinberg \cite{Steinberg:1976}.
The initial phase of node $k$ 
in \eqref{varphik1} 
will now be given by
\begin{equation}
\hat{\Psi}_k = - \frac{2\pi}{\lambda} d_k(\phi_0, \theta_0)
+ \varphi_k
\end{equation}
where $\varphi_k$ corresponds to the phase offset
due to the phase ambiguity caused by carrier phase jitter
or offset between the transmitter and receiver nodes.
In the following, the phase offset $\varphi_k$'s are 
assumed to be i.i.d. random variables.
Then, from \eqref{far_field_F0}, \eqref{Dk_0}, \eqref{far_field_F1},
and \eqref{far_field_F4},
the far-field array factor 
(with $\theta = \theta_0=\pi/2$) will be given by
\begin{align}
\tilde{F}(\phi | {\boldsymbol z} , {\boldsymbol \varphi} ) 
& =
\frac{1}{N} \sum_{k=1}^{N} e^{ j 
\left( - z_k
{4 \pi} \tilde{R}
\sin \frac{\phi}{2} 
  + \varphi_k \right) } 
=
\frac{1}{N} \sum_{k=1}^{N} e^{ - j 
z_k {4 \pi} \tilde{R}
\sin \frac{ \phi}{2} 
 } e^{j  \varphi_k } .
\label{F_with_phase1}
\end{align}
The average beampattern of \eqref{av_beam0} will be replaced by
\begin{align}
P_{\text{av}}( \phi)
& \triangleq E_{{\boldsymbol z},  { \boldsymbol \varphi}}
\left\{
P( \phi | {\boldsymbol z},  {\boldsymbol \varphi} )  
\right\} .
\label{F_with_phase2}
\end{align}
Similar to \eqref{av_beam1},
direct calculation of 
\eqref{F_with_phase2} results in
\begin{align}
P_{\text{av}}( \phi )
& = \frac{1}{N} + \left( 1 - \frac{1}{N} \right)
\left| 2 \frac{J_1\left(
\alpha(\phi)
\right)}
{ \alpha(\phi) }
\right|^2 
\left|
A_{\varphi}
\right|^2
\end{align}
where
\begin{equation}
A_{\varphi} \triangleq 
E_{ \varphi_k} \left\{
 e^{j \varphi_k} \right\} .
\end{equation}
Thus, as $N \to \infty$,
the average beampattern will simply become a 
version of the original scaled by a factor of $|A_{\varphi}|^2$.

\begin{figure}[t]
\begin{center}
\includegraphics[width=12.5cm,clip]{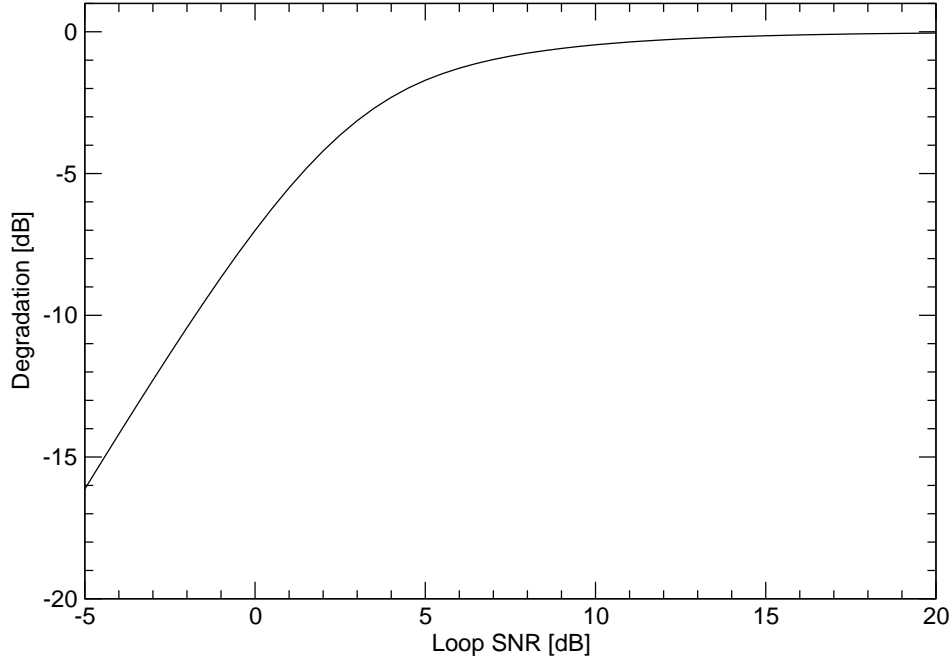}
\caption{Mainbeam degradation due to the phase noise in the closed-loop scenario.}
\label{fig:degradation_close}
\end{center}
\vspace{-1cm}
\end{figure}

Let us now assume that the phase offset follows a Tikhonov distribution,
a typical phase jitter model for phase-locked loop (PLL) circuits
given in \cite{Viterbi:1966},
\begin{equation}
f_{\varphi}(x) = \frac{1}{2\pi I_0\left(1/\sigma^2_{\varphi}\right)}
\exp\left( \cos(x) / \sigma^2_{\varphi} \right), \qquad |x| \leq \pi ,
\end{equation}
where $\sigma^2_{\varphi}$ is the variance of the phase noise and
$I_n$ is the $n$th order modified Bessel function of the first kind.
The corresponding attenuation factor is given by
\begin{equation}
A_{\varphi} = \frac{I_1(1/\sigma^2_{\varphi})}
{I_0(1/\sigma^2_{\varphi})} .
\end{equation}
The variance of the phase noise $\sigma^2_{\varphi}$
is related to the loop SNR of the PLL by
\begin{equation}
\rho_{\varphi} = 1/ \sigma^2_{\varphi}  .
\end{equation}
Fig. \ref{fig:degradation_close} shows the degradation 
factor  $|A_\varphi|^2$ with respect to the loop SNR.
As observed from the figure, a loop SNR of at least 3\,dB
may be necessary
for each node in order to reduce the overall beampattern 
degradation to less than 3\,dB.

\subsection{Open-loop Case}

In the open-loop case, our model of the initial phase is given
in \eqref{varphik2} with $\theta_0 = \frac{\pi}{2}$, and if there are estimation errors
in the location parameters $r_k$ and $\psi_k$, 
the initial phase will be replaced by 
\begin{align}
\hat{\Psi}^{\dag}_k & = 
\frac{2\pi}{\lambda}
({r}_k + \delta {r}_k)
\cos(\phi_0 - ({\psi}_k  + \delta \psi_k) ) 
\nonumber \\
& =
\frac{2\pi}{\lambda}
{r}_k 
\cos(\phi_0 - ({\psi}_k  + \delta \psi_k) )
+
\frac{2\pi}{\lambda}
\delta {r}_k
\cos(\phi_0 - ({\psi}_k  + \delta \psi_k) ),
\end{align}
where $\delta r_k$ and $\delta \psi_k$ are the corresponding error
random variables, each set assumed to be i.i.d. and also independent
of $r_k$ and $\psi_k$ for simplicity.
With the far-field approximation, we have
\begin{align}
\frac{2\pi}{\lambda} d_k \left(\phi, \frac{\pi}{2} \right)
+ \hat{\Psi}_k^{\dag}
& \approx
\frac{2 \pi}{\lambda}  \left\{
A
- 
{r}_k \left[
\cos(\phi - \psi_k)
- 
\cos(\phi_0 - {\psi}_k  - \delta \psi_k )
\right]
 +
\delta {r}_k
\cos(\phi_0 - ({\psi}_k  + \delta \psi_k) )
\right\}
\nonumber \\
& = 
\frac{2\pi}{\lambda} A + 
\frac{4 \pi}{\lambda} {r}_k
\left[
\sin\left( \psi_k - 
\frac{\phi_0 + \phi - \delta \psi_k}{2}
\right)
\sin\left(
\frac{\phi_0 - \phi - \delta \psi_k}{2}
\right)
\right] \nonumber \\
& \quad +
\frac{2\pi}{\lambda}
\delta {r}_k 
\cos( {\psi}_k   - 
(\phi_0 - \delta {\psi}_k) )  .
\label{dist_phase2}
\end{align}
Let $\tilde{\psi}_k \triangleq \psi_k
- \frac{  \phi + \phi_0 - \delta \psi_k}{2}$.
Then, the right-hand side (RHS) of 
\eqref{dist_phase2} is given by
\begin{align}
& 
\frac{2\pi}{\lambda}
A - 
\frac{4 \pi}{\lambda} {r}_k
\sin \tilde{\psi}_k 
\sin\left(
\frac{ \phi - \phi_0- \delta \psi_k}{2}
\right)
+ \frac{2\pi}{\lambda}
\delta {r}_k 
\cos \left( \tilde{\psi}_k   + 
\frac{  \phi - \phi_0 + \delta \psi_k}{2}
\right) .
\end{align}
The resulting far-field array factor
of \eqref{far_field_F2} will then be given by
\begin{align}
\tilde{F}^{\dag} (\phi | {\boldsymbol r},{\boldsymbol \psi},
{\boldsymbol \delta
\boldsymbol \psi},  
{\boldsymbol \delta
\boldsymbol r} ) 
& =e^{j \frac{2 \pi}{\lambda}
A}
\frac{1}{N}  \sum_{k=1}^{N} 
 e^{- j \frac{4 \pi} {\lambda}
{r}_k \sin \tilde{\psi}_k 
\sin\left(
\frac{  \phi - \phi_0 - \delta \psi_k}{2}
\right) +j 
\frac{2\pi}{\lambda}
\delta{r}_k 
\cos \left( \tilde{\psi}_k  + 
\frac{ \phi - \phi_0 + \delta \psi_k}{2}
\right) },
\end{align}
and the beampattern is expressed as
\begin{align}
{P} ( \phi| {\boldsymbol z},{\boldsymbol v},
{\boldsymbol \delta
\boldsymbol \psi} ) 
& = \frac{1}{N}+\frac{1}{N^2}\sum_{k=1}^{N} 
\sum_{\stackrel{ \scriptstyle l=1}{l\neq k}}^{N} 
 e^{-j{4 \pi} \tilde{R} 
\left\{
z_k \sin\left(
\frac{  \phi - \phi_0 - \delta \psi_k   }{2}
\right)
-
z_l \sin\left(
\frac{   \phi - \phi_0 - \delta \psi_l  }{2}
\right)
\right\}}
 e^{j \frac{2\pi}{\lambda}
\left( v_k - v_l \right)
},
\end{align}
where 
\begin{align}
z_k & \triangleq \frac{{r}_k}{R} \sin \tilde{\psi}_k 
= \tilde{r}_k \sin \left(
\psi_k +
\frac{\delta \psi_k}{2} - 
\frac{ \phi + \phi_0 }{2}  \right)
 \\
v_k & \triangleq \delta {r}_k 
\cos \left( \tilde{\psi}_k  + 
\frac{  \phi + \delta \psi_k}{2}
\right) 
= \delta {r}_k 
\cos \left( \psi_k  + 
\delta \psi_k - \phi_0
\right) .
\end{align}
Conditioned on $\phi, \phi_0$ and $\delta \psi_k$, 
the angle $\tilde{\psi}_k$ can be seen as a uniformly distributed
random variable, and thus the pdf of $z_k$ is given by \eqref{pdf_z}.
Considering the fact that $r_k$ and $\delta r_k$ are 
assumed to be statistically independent, we further
assume for analytical purposes that $z_k$ and $v_k$ are
statistically independent. Then, again, the beampattern does not depend
on the particular choice of $\phi_0$.
Furthermore, on modeling $\delta {r}_k$ as being uniformly distributed
over $[-r_{\text{max}}, r_{\text{max}}]$ and assuming the phase term of
$v_k$ to be uniformly distributed over $[0, 2 \pi]$, 
the probability density function of $v_k$ will be given by
\begin{equation}
f_{v_k} (v) = \frac{1}{\pi r_{\text{max}}} \left[
\ln \left( 1 + \sqrt{1 - \left(\frac{v}{r_{\text{max}}}\right)^2}\right)
- \ln \frac{|v|}{r_{\text{max}}}
\right], \qquad |v| \leq r_{\text{max}} .
\end{equation}
Consequently, the average beampattern can be written as
\begin{align}
{P}_{\text{av}} (\phi ) 
& = \frac{1}{N}+\left( 1 - \frac{1}{N}\right)
\left| A_{\psi} (\phi) \right|^2
\left| A_{r} \right|^2 ,
\end{align}
where
\begin{align}
A_{r} & \triangleq
E_{v_k} \left\{
 e^{j \frac{2\pi}{\lambda} v_k  }
\right\}  =  \frac{2}{\pi}
\int_{0}^{1}
\cos\left( \frac{2 \pi}{\lambda}  r_{\text{max}} t \right) 
\ln \frac{ 1 + \sqrt{1 - t^2} }{t}
dt  \nonumber \\
& = 
 {\,}_1 F_2
\left( \frac{1}{2} \,;\, 1, \frac{3}{2} \,;\,
- \left( \pi \frac{r_{\text{max}}}{\lambda} \right)^2
\right)
\label{Ar} \\
 A_{\psi} (\phi) & \triangleq
E_{z_k, \delta \psi_k} \left\{
 e^{j{4 \pi} \tilde{R} 
z_k \sin\left(
\frac{\phi_0 + \delta \psi_k - \phi }{2}
\right)
}
\right\}
 = 
E_{\delta \psi_k}
\left\{
\frac{J_1\left(
{4 \pi} \tilde{R}
\sin  \frac{  \phi - \delta \psi_k }{2}  \right)}
{{2 \pi} \tilde{R}
\sin  \frac{ \phi -\delta \psi_k }{2} }
\right\},
\label{Apsi}
\end{align}
and without loss of generality $\phi_0 = 0$ was assumed.
In \eqref{Ar}, 
${\,}_1 F_2 \left( \frac{1}{2} \,;\, 1, \frac{3}{2} \,;\,
- x^2 \right)$ denotes a generalized hypergeometric function 
which has an oscillatory tail but 
converges to zero as $x$ increases.

Also, assuming that the $\delta \psi_k$ are
uniformly distributed over $[- \psi_{\text{max}}, \psi_{\text{max}}]$
and using the approximation
$\sin \left(  \phi + \delta \psi_k \right)
\approx \phi + \delta \psi_k$
which is valid for the beampattern around the mainbeam,  
we obtain
\begin{align}
 A_{\psi} (\phi)  
& \approx
\frac{1}{2}
\left(
1 - \frac{ \phi }{\psi_{\text{max}}}
\right) 
 {\,}_1 F_2
\left( \frac{1}{2} \,;\, \frac{3}{2}, 2 \,;\,
- ( \pi \tilde{R} ( \phi + \psi_{\text{max}}  ) )^2
\right)
\nonumber \\
& \quad 
+ 
\frac{1}{2}
\left( 1
 + \frac{ \phi }{\psi_{\text{max}}}
\right) 
 {\,}_1 F_2
\left( \frac{1}{2} \,;\, \frac{3}{2}, 2 \,;\,
- ( \pi \tilde{R} ( \phi - \psi_{\text{max}} ) )^2
\right)  .
\end{align}
Since the hypergeometric function
${\,}_1 F_2
\left( \frac{1}{2} \,;\, \frac{3}{2}, 2 \,;\,
-  x^2 \right)$ has a maximum peak value of 1 at $x = 0$,
the above expression indicates that 
regardless of the value of
$\tilde{R}$, there may be two symmetric peaks around the mainbeam
at $ \phi = \pm \psi_{\text{max}}$ resulting in a {\em pointing error}.
Therefore, the mainbeam may spread over by a factor 
of $\psi_{\text{max}}$.
At the center of the mainbeam, we have
\begin{align}
A_{\psi} (0) & =
 {\,}_1 F_2
\left( \frac{1}{2} \,;\, \frac{3}{2}, 2 \,;\,
- \left( \pi \frac{R\psi_{\text{max}}}
{\lambda} 
\right)^2
\right)  .
\end{align}

\begin{figure}[t]
\begin{center}
\psfrag{rmax/l or Rpmax/l}[Bl][Bl][.8]{$r_{\text{max}}/\lambda$ \text{\sf
 or } ${R \psi_{\text{max}} / \lambda}$}
\includegraphics[width=12.5cm,clip]{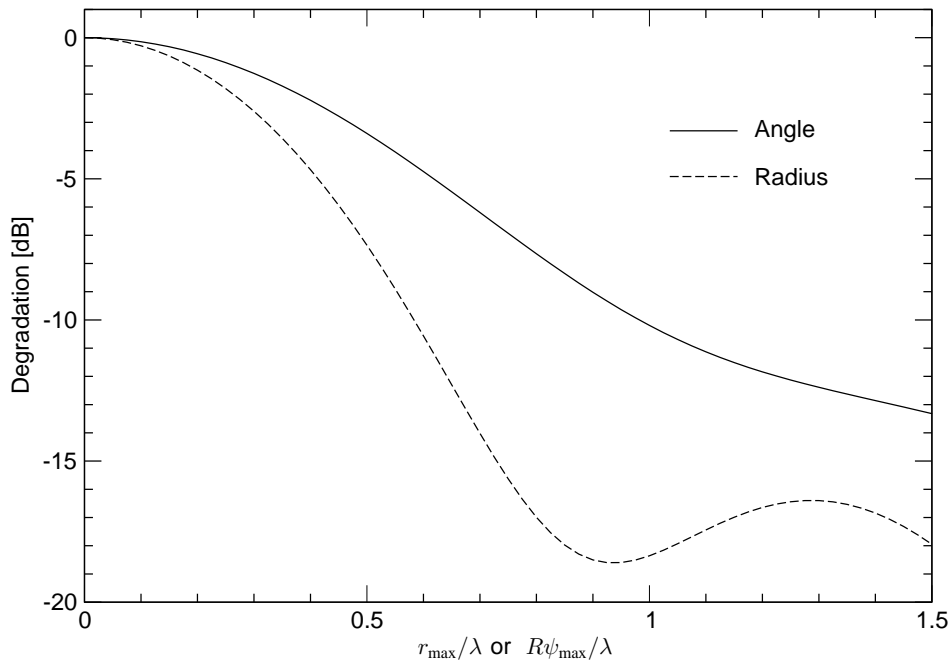}
\caption{Mainbeam degradation due to location estimation errors in the open-loop scenario.}
\label{fig:degradation_open}
\end{center}
\vspace{-1cm}
\end{figure}

Fig. \ref{fig:degradation_open} shows the degradation 
factor  $|A_r|^2$  and  $|A_{\psi} (0)|^2$ 
for a given $\frac{r_{\text{max}}}{\lambda}$ and 
$\frac{R \psi_{\text{max}}}{\lambda}$.
As observed from the figure and discussion above,
the angle estimation error has two effects, i.e., pointing
error and mainbeam degradation. In particular, if we wish to suppress
the mainbeam degradation below 3\,dB, from the figure, we should 
choose $R\psi_{\text{max}}/ \lambda \leq 1/2$. This means that
the maximum angle estimation error should satisfy
\begin{equation}
\psi_{\text{max}} \leq \frac{\lambda}{2R} = \frac{1}{2\tilde{R}},
\label{cond_min_angle}
\end{equation}
and as $\tilde{R}$ becomes large, the requirement
of minimum angle ambiguity from \eqref{cond_min_angle} becomes
severe.

\section{Conclusion}
\label{sec:conclusion}

In this paper, we have analyzed the stochastic performance 
of random arrays for distributed collaborative beamforming, 
in the framework of wireless
ad hoc sensor networks. It has been shown that under ideal channel 
and system assumptions, directivity of order $N$ can be 
achieved asymptotically with $N$ sensor nodes,
as long as the sensor nodes are located sparsely enough. We 
have studied
the average and the distribution of the beampattern as well as
the distribution of the sidelobe peaks. 
Several forms of the CCDF of the beampattern have been derived
and compared, with particular emphasis on the Gaussian approximation
of the array factor. 
We have considered two scenarios of distributed
beamforming and investigated
the effects of phase ambiguity and location
estimation error upon the resultant average beampatterns.

Our main conclusion is that, given a number of nodes
randomly distributed over a large disk, one may form a nice
beampattern with narrow mainlobe and sidelobes as low as $1/N$ plus
some margin for maximum sidelobe peaks.
Also, the directivity approaches $N$ if the nodes are located
as sparsely  as possible. 
However, our analysis is based on a number of
ideal assumptions on the system and channel model. 
In practice, a number of open issues
remain, such as applicability of beamforming 
when the destination or nodes in the cluster 
are in rapid motion or the channel suffers severe multipath fading.
Also, specific algorithms should be developed for frequency offset
correction of each node as well as methods for initial phase or 
location estimation. Finally, 
efficient protocols for sharing the transmit as well as calibration
information among nodes are required.

\appendices

\section{Proof of Theorem \ref{theorem_directivity}}
\label{app:directive}

We first prove the following lemma:
\begin{lemma} 
\label{lemma_hyper}
A generalized hypergeometric function 
${\,}_2 F_3 
\left( \frac{1}{2}, \frac{3}{2} \,;\, 1, 2, 3 \,;\,
- x^2 \right)$ 
with $x \gg 1$ can be bounded as
\begin{equation}
f(x) \triangleq
 {\,}_2 F_3 
\left( \frac{1}{2}, \frac{3}{2} \,;\, 1, 2, 3 \,;\,
- x^2 \right)
\leq \frac{c_0}{x}
\end{equation}
where $c_0$ is a constant ($c_0 \approx 1.1727$).
\end{lemma}

\begin{proof}
We start with the integral form
\begin{align}
f(x) & =
\frac{1}{\pi} \int_0^{\pi} 
\left|
2 \frac{J_1 \left(x \sin \frac{\theta}{2} \right)}
{x \sin \frac{\theta}{2} }
\right|^2  d \theta
 = 
\frac{1}{\pi} \int_0^{x}
\left|
2 \frac{J_1 \left( t \right)}
{ t }
\right|^2 
 \frac{2}{\sqrt{x^2 - t^2}}
 d t .
\label{fx}
\end{align}
Since the asymptotic form of $J_1(x) $ given by
\eqref{bessel_asympto} is valid
for $x \gg 1$, 
we have the following inequalities
\begin{align}
\left|
2 \frac{J_1 \left( t \right)}
{ t }
\right|^2 
 & \leq  \cos^2( \alpha_0 t ), \qquad \text{ for }  t \leq x_0 
\label{relation_J1}
\end{align}
\begin{align}
\int_t^{t+\Delta}
\left|
2 \frac{J_1 \left( u \right)}
{ u } 
\right|^2  \,du
 & \leq   \int_t^{t+\Delta}
\frac{8}{\pi u^3}
\,du,
\qquad \text{ for }  t > x_0 
\label{relation_J2}
\end{align}
for some threshold value $x_0$
which should be determined numerically, and  
for some interval $\Delta > 0$.
The parameter $\alpha_0$ is chosen to be the smallest
non-negative value such that 
\begin{equation}
\cos(\alpha_0 x_0) = \sqrt{\frac{8}{\pi x^3_0}},
\end{equation}
should hold, and this guarantees a 
continuity of the function at the threshold $t = x_0$.
Fig.\,\ref{fig:function_approximation} illustrates the relationship
of \eqref{relation_J1} and \eqref{relation_J2}  
with $x_0$ chosen as a cross point of $t$
between the functions $J_1(t)$ and $\sqrt{2 / (\pi t)}$,
yielding $x_0 = 2.4445$. The corresponding value
of $\alpha_0$ is 0.4664.
\begin{figure}[t]
\begin{center}
\psfrag{Function of t}[Bl][Bl][1.0]{{\sf Function of }$t$}
\psfrag{t}[Bl][Bl][1.0]{$t$}
\psfrag{bessel}[Bl][Bl][.8]{$|2J_1(t)/t |^2$}
\psfrag{asymptote}[Bl][Bl][.8]{$8/(\pi t^3)$}
\psfrag{cosf}[Bl][Bl][.8]{$\cos^2(\alpha_0 t)$}
\includegraphics[width=12.5cm,clip]{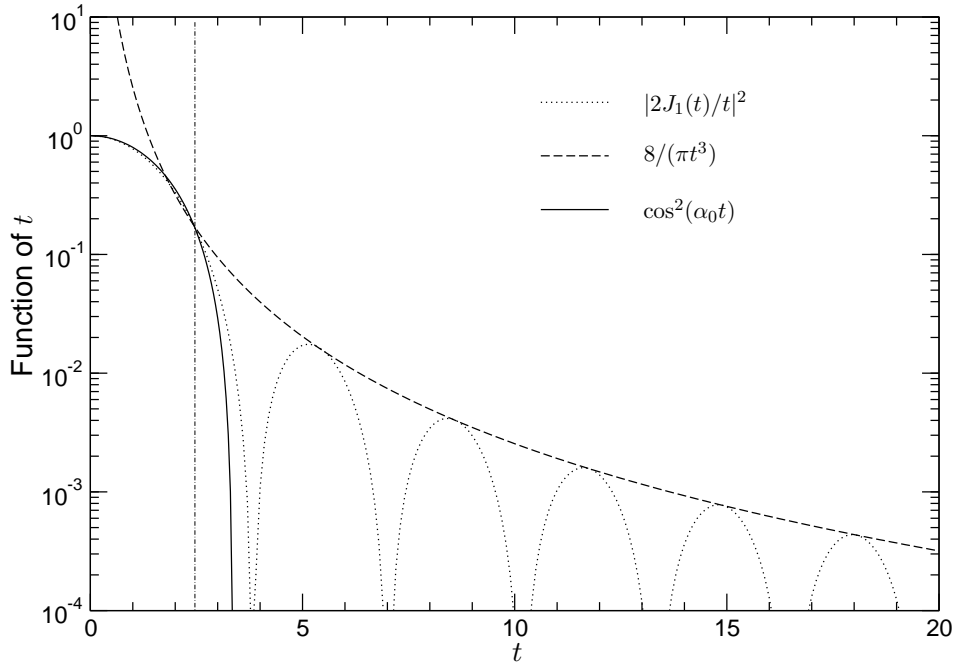}
\caption{Function $|2J_1(t)/t |^2$ and its upper bound with $x_0=2.4445$.}
\label{fig:function_approximation}
\end{center}
\vspace{-1cm}
\end{figure}
Substituting \eqref{relation_J1} and \eqref{relation_J2} 
into \eqref{fx}, we get for $x > x_0$
\begin{align}
f(x) & \leq
\frac{2}{\pi} \int_0^{x_0}
 \frac{\cos^2 \left(
\alpha_0 t
\right)
}{\sqrt{x^2 - t^2}}
 d t 
+
\frac{16}{\pi^2} \int_{x_0}^{x}
 \frac{1}{ t^3 \sqrt{x^2 - t^2}}
 d t .
\label{fx2}
\end{align}
The first term on the RHS of \eqref{fx2}
is given by
\begin{align}
\frac{2}{\pi} \int_0^{x_0}
 \frac{\cos^2 \left(
\alpha_0 t
\right)
}{x \sqrt{1 - \left(
\frac{t}{x}\right)^2}}
 d t 
& =
\frac{2}{\pi x} \int_0^{x_0}
\left\{
1 + 
\frac{1}{2}
\left(
\frac{t}{x}\right)^2
+ O\left(
1/x^4
\right)
\right\}
{\cos^2 \left(
\alpha_0 t
\right)
} d t 
\nonumber \\
& =
\frac{1}{\pi x} 
\left(
{x_0}
+
\frac{ \sin(2\alpha_0 x_0)}{2 \alpha_0}
\right)
+ O\left(
1/x^3
\right)  .
\label{fx3_1}
\end{align}
The second term on the RHS of \eqref{fx2}
is given by
\begin{align}
\frac{16}{\pi^2} \int_{x_0}^{x}
 \frac{1}{ t^3 \sqrt{x^2 - t^2}}
 d t
& = 
\frac{16}{\pi^2} 
\left[
\frac{1}{x}
\frac{ \sqrt{1 - \left( \frac{x_0}{x} \right)^2 }}
{2 x^2_0} 
+ \frac{1}{2 x^3} \left\{
\ln\left( 1 + \sqrt{1 - \left(\frac{x_0}{x} \right)^2} \right) 
+ \ln \left( \frac{x}{x_0} \right)
\right\}
\right]
\nonumber \\
& = \frac{8}{\pi^2 x^2_0}
\frac{1}{x}
\left(
1 -
\left(
\frac{x_0}{x}
\right)^2 + O\left(1/x^4 \right)
\right)
+ O\left( \ln(x)/x^3 \right)
\nonumber \\
& = \frac{1}{x} \frac{8}{\pi^2 x^2_0}
+ O\left( \ln(x)/x^3 \right) .
\label{fx3_2}
\end{align}
Consequently, we may write
\begin{align}
f(x) & \leq 
\frac{1}{\pi}
\left(
{x_0}
+
\frac{ \sin(2\alpha_0 x_0)}{2 \alpha_0}
+
 \frac{8}{\pi x^2_0}
\right)
\frac{1}{x} + O\left( \ln(x) / x^3 \right)
\label{fx_final}
\end{align}
and the second term on the RHS of \eqref{fx_final} drops as $x$ becomes
large. 
With $x_0 = 2.4445$ and $\alpha_0 = 0.4664$, the coefficient of $1/x$
can be calculated to be $c_0 = 1.1727$.
\end{proof}

\begin{proof}[Proof of Theorem \ref{theorem_directivity}]
From \eqref{jensen_relation}, \eqref{Dav2}, 
 and Lemma \ref{lemma_hyper}, we have
\begin{align}
\frac{{D}_\text{av}}{N} & \geq 
\frac{\tilde{D}_\text{av}}{N} \geq 
\frac{1}{1 + (N-1) \frac{c_0}{4 \pi \tilde{R}}}
=
\frac{1}{1 + \left( 1 - \frac{1}{N} \right) \frac{c_0}{4 \pi}
\frac{N}{\tilde{R}}}  .
\label{final_theorem}
\end{align}
For large $N$, the RHS of \eqref{final_theorem} converges
to \eqref{theorem1_eq} with $\mu = \frac{c_0}{4\pi} \approx 0.09332$.
\end{proof}

\section{The Mean Number of Upward Level Crossings of a Gaussian Process}
\label{app:level_crossing}

In this appendix, we obtain the mean number of upward 
crossings of a given level of the zero mean Gaussian process
based on the approach of Rice \cite{Rice:1944,Rice:1945}.
Assume that ${X}$ and ${Y}$ are uncorrelated
zero-mean Gaussian processes with variance $\sigma^2_x = \sigma^2_y = 1/2$.
Let $u = \sin \left(\frac{\phi}{2} \right)$ and
${X}'$ and ${Y}'$ denote
the corresponding processes differentiated by $u$.
By assumption, ${X}'$ and ${Y}'$ become
zero mean Gaussian processes. 
In order to calculate the variance, first consider the 
autocorrelation function of ${X}$
at instants $u = u_1$ and $u_2$ given by
\begin{align}
\rho_{{X}}(u_1, u_2) & = 
E_{z} \left\{
\cos\left( z 4 \pi \tilde{R}  u_1   \right)
\cos\left( z 4 \pi \tilde{R} u_2  \right)
\right\} + \text{other terms}
\nonumber \\
& = \frac{1}{2} E_{z} \left\{
\cos\left( z 4 \pi \tilde{R}  \left( u_1 + u_2 \right)   \right)
\right\}
+  \frac{1}{2} E_{z} \left\{
\cos\left( z 4 \pi \tilde{R}  \left( u_1 - u_2 \right)   \right)
\right\}, 
\label{rho_first}
\end{align}
where the other terms become zero by the zero mean assumption.
Also for the same reason, the first term of the RHS of
\eqref{rho_first} may be approximated by zero. Therefore,
letting $v=u_1 - u_2$, we obtain
\begin{align}
\rho_{{X}}(v) & \approx
 \frac{1}{2} E_{z} \left\{
\cos\left( z 4 \pi \tilde{R}  v   \right)
\right\} .
\label{rho_second}
\end{align}
Differentiating the above with respect to $v$ twice,
setting $v=0$, and carrying out the statistical average with respect to $z$, 
the variance of ${X}'$ is given by \cite{Cramer:1967}
\begin{equation}
\sigma^2_{{x}'} =  -  \rho''_{{X}}(0)
 =  2 \pi^2 \tilde{R}^2 .
\end{equation} 
Likewise, one may obtain $\sigma^2_{{y}'} = \sigma^2_{{x}'}$, and 
the joint pdf of ${X}, {X}', {Y}, {Y}'$
is given by
\begin{equation}
f_{{X}, {Y}, {X}', {Y}'}
(x, y, x', y') = 
\frac{1}{(2\pi)^2 \sigma^2_{x} \sigma^2_{{x}'} }
\exp \left(
- \frac{x^2 + y^2}{2\sigma^2_{x}}
- \frac{x'^2 + y'^2}{2\sigma^2_{x'}}
\right) .
\end{equation}
On changing the random variables in the polar coordinates
via ${X}= \Omega \cos \Theta$, ${Y} = 
\Omega \sin \Theta$ and integrating out
$\Theta$ and $\Theta'$, we obtain
\begin{equation}
f_{\Omega, \Omega'}(\omega, \omega')
 = 
 \omega e^{-\omega^2}
\frac{1}{\sqrt{\pi} \pi \tilde{R}} e^{-\frac{\omega'^2}{4 \pi^2 \tilde{R}^2}} .
\end{equation}
The number of positive (upward) crossings of the process $\omega$
at level $a$ per interval $du$
is given by \cite{Rice:1944,Rice:1945}
\begin{equation}
\nu(a) du  = du
\int_0^{\infty} \omega' f_{\Omega, \Omega'}
(a, \omega' ) d\omega'
= du  \, 2 \sqrt{\pi} \tilde{R} a e^{-a^2} .
\label{mean_crossing}
\end{equation}
Consequently, the mean number of upward crossings 
for the interval ${\cal S}_{3\text{dB}}$
is given by
\begin{equation}
E \left\{
\nu(a) 
\right\}
= 
\int_{u = \sin \left( \frac{\phi}{2} \right),
\phi \in {\cal S}_{3\text{dB}}}
du \,\,
\nu(a) ,
\end{equation}
which results in \eqref{mean_crossing2}.

\bibliographystyle{ieeetr}

\end{document}